\documentclass[a4paper,11pt]{article}
\usepackage[sumlimits,intlimits,namelimits]{amsmath}
\usepackage{amssymb}
\usepackage[english]{babel}
\usepackage{bbm}
\usepackage{parskip,wasysym}

\usepackage{graphicx}          

\usepackage[usenames,dvipsnames]{color}

\newcommand{\mA}{\mathcal{A}}
\newcommand{\mB}{\mathcal{B}}
\newcommand{\mC}{\mathcal{C}}
\newcommand{\mD}{\mathcal{D}}
\newcommand{\mE}{\mathcal{E}}
\newcommand{\mF}{\mathcal{F}}
\newcommand{\mM}{\mathcal{M}}
\newcommand{\mS}{\mathcal{S}}
\newcommand{\mL}{\mathcal{L}}

\newcommand{\mR}{\mathcal{R}}

\global\parskip 6pt

\newcommand{\qed}{\nobreak \ifvmode \relax \else\ifdim\lastskip<1.5em \hskip-\lastskip\hskip1.5em plus0em minus0.5em \fi \nobreak\vrule height0.75em width0.5em depth0.25em\fi}

\newcommand{\be}{\begin{eqnarray}}
\newcommand{\ee}{\end{eqnarray}}

\def\>{\rangle}
\def\<{\langle}

\usepackage{jheppub} 

\usepackage[T1]{fontenc} 

\title{\boldmath Conditions on holographic entangling surfaces in higher curvature
 gravity}
\preprint{MPP-2014-8}

 \author{Johanna Erdmenger,}
 \author{Mario Flory}
  \author{and Charlotte Sleight}
 \affiliation{Max-Planck-Institut f\"ur Physik (Werner-Heisenberg-Institut),\\F\"ohringer Ring 6, D-80805, Munich, Germany}

\emailAdd{jke@mpp.mpg.de}
\emailAdd{mflory@mpp.mpg.de}
\emailAdd{csleight@mpp.mpg.de}

\abstract{We study the extremal surfaces of functionals recently proposed for the 
holographic calculation of entanglement entropy in general higher curvature 
theories, using New Massive gravity and Gauss-Bonnet gravity as concrete 
examples. We show that the entropy functionals admit closed extremal surfaces, 
which for black hole backgrounds can encircle the event horizon of the black 
hole. In the examples considered, such closed surfaces correspond to a 
lower 
value of the entropy functional than expected from CFT calculations, 
implying a seeming mismatch between the bulk and boundary calculations. 
For Lorentzian settings we show that this problem can be 
resolved by imposing a causality constraint on the extremal 
surfaces. The possibility of deriving conditions from an alternative 
conical boundary condition method as proposed by Lewkowycz and Maldacena is 
explored.}

\begin{document} 
\maketitle
\flushbottom

\section{Introduction}
\label{sec::intro}
\numberwithin{equation}{section}
The concept of entanglement entropy has proven to be of great interest to the 
physics community in recent years. The search of a better understanding of 
quantum systems, in particular at criticality, has lead to the establishment of 
general analytic results for entanglement entropy in two-dimensional CFTs (see 
e.g. \cite{Calabrese:2004eu}), while in higher dimensions the study of 
entanglement entropy is less simple. Entanglement entropy also appears to be 
instrumental for a deeper understanding of holography 
\cite{PhysRevD.86.065007,VanRaamsdonk:2009ar}.

The development of an understanding of entanglement entropy in the framework 
of the AdS/CFT correspondence 
\cite{Maldacena:1997re,Gubser:1998bc,Witten:1998qj} is therefore of utmost 
interest. A simple example of the AdS/CFT correspondence at play in this context 
is given by the Bekenstein-Hawking entropy 
\cite{Bekenstein:1973ur,Hawking:1974sw}
\begin{equation}
S_{BH} = \frac{\text{Area of Horizon}}{4 G_{N}},
\end{equation}
which for BPS black holes has a microscopic derivation in string theory 
\cite{Strominger:1996sh}, implying a relation between gravitational entropy and 
the degeneracy of quantum field theory as its microscopic description. 
Important progress was made by Ryu and Takayanagi (RT) 
\cite{Ryu:2006bv,Ryu:2006ef}, who generalised the Bekenstein-Hawking formula 
above to the holographic calculation of entanglement entropy in CFTs with 
Einstein-Hilbert gravity duals. Here, the entanglement entropy is given by the 
area of a minimal surface which is anchored to the given spatial region in the 
CFT. This prescription for entanglement entropy was recently proven by Lewkowycz 
and Maldacena in \cite{Lewkowycz:2013nqa} by translating the replica trick 
to the classical gravity action in the bulk, and showing that the resultant 
prescription for holographic entanglement entropy is equivalent to that of RT. 
However to date there is no proof of its covariant generalisation by Hubeny, 
Rangamani and Takayanagi (HRT) \cite{Hubeny:2007xt}.

More recently, generalisations of the RT and HRT formulae to general higher 
derivative theories of gravity have been proposed 
\cite{Fursaev:2013fta,Miao:2013nfa,Dong:2013qoa,Camps:2013zua}. 
In black hole backgrounds, for consistency a correct prescription should reduce 
to Wald's entropy functional 
\cite{Wald:1993nt,Jacobson:1993vj,Iyer:1994ys}

\begin{equation}
S_{\text{Wald}} = -2 \pi \int_{\mB} d^{d-1}y \sqrt{\gamma} \frac{\partial 
L}{\partial R_{\mu \rho \nu \sigma}} \varepsilon_{\mu \rho} \varepsilon_{\nu 
\sigma}, \label{wald}
\end{equation}

when evaluated on the black hole bifurcation surface $\mB$ (see appendix 
\ref{sec::notation} for notation). Despite parallels to the 
Einstein-Hilbert case, Wald's functional does not provide a prescription for the 
calculation of entanglement entropy in field theory duals to higher curvature 
gravity \cite{Hung:2011xb}, and the current proposals take into account 
the extrinsic curvature of the surfaces upon which they are evaluated.

When using functionals to calculate entanglement entropy, they must be 
evaluated on a particular surface in the bulk anchored to the region of interest 
in the boundary CFT. One way to locate these surfaces could be to directly 
extremise the functionals, which has already been proposed in 
\cite{Dong:2013qoa}. In this paper we investigate this method for static and partially 
static backgrounds (i.e. AdS and black hole backgrounds) in New Massive 
Gravity (NMG) and Gauss-Bonnet gravity, with a particular focus on the nature of the closed extremal surfaces of the 
corresponding 
functionals. In black hole backgrounds, a particularly important closed 
extremal surface is the black hole bifurcation surface. When evaluated on the 
black hole bifurcation surface, the entanglement entropy functional yields 
Wald's formula \eqref{wald} for black hole entropy. This coincides with 
the dual CFT entropy by standard AdS/CFT lore \cite{Witten:1998zw}.

We find there exists an extra closed extremal surface for certain parameter 
ranges in NMG and Gauss-Bonnet gravity. Such surfaces can give a lower value 
for the entropy than that expected from CFT calculations. A prescription for 
calculating entanglement entropy based on only extremising 
the functionals would hence lead us to incorrectly equate the dual CFT entropy 
with that given by the additional closed extremal surface. Our findings lead 
us to conclude that, for consistency and physical values for entanglement 
entropy, additional constraints on the extremal surfaces need to be imposed. 
In section \ref{sec::Headrick} we show that in a covariant setting such 
constraints can be motivated by causality. For Euclidean settings, 
since causality cannot be used to derive constraints on the surfaces, in 
section \ref{sec::GBconstraints} we investigate whether constraints can be 
derived by solving the equations of motion with conical boundary conditions.

\subsection*{Summary of results}
The main findings of our paper can be summarised as follows: The 
functionals proposed for the calculation of holographic entanglement entropy in 
higher curvature gravities allow for extremal surfaces in addition to those 
that correspond to the correct physical value of entanglement entropy. We find 
that existing prescriptions for the holographic calculation of entanglement 
entropy are not sufficient, as in some cases they would lead one to incorrectly 
pick an unphysical additional surface as the one that is supposed to yield the 
CFT result. We examine examples of such additional surfaces in NMG and 
Gauss-Bonnet gravity. For Lorentzian settings we present an argument based 
on 
causality that can be employed to consistently eliminate the additional surfaces 
found. For Euclidean settings, we investigate the possibility of deriving 
conditions that entangling surfaces must satisfy by considering the extension of 
the replica trick into the bulk. The arguments we present for both Lorentzian 
and Euclidean settings can in principle be applied to any gravitational theory 
that admits asymptotically AdS solutions.
  
This paper is organised as follows: In section \ref{sec::2} we introduce static 
and covariant prescriptions for calculating entanglement entropy in both the 
Einstein-Hilbert and higher curvature gravity theories. For higher curvature 
gravity theories, we define these as the RT-like and HRT-like prescriptions, 
respectively. In section \ref{sec::closedcurves} we explain the significance of 
the closed extremal surfaces of holographic entanglement entropy functionals, in 
particular the role played by the homology constraint in the interpretation of 
such surfaces. In sections \ref{sec::NMGtotal} and \ref{sec::GBG} we study the 
nature of the extremal surfaces of the entropy functionals in NMG and 
Gauss-Bonnet gravity, paying particular attention to their closed extremal 
surfaces in the bulk. We explain that in using the functionals to calculate 
entanglement entropy, constraints on the extremal surfaces additional to the 
homology constraint need to be imposed. In section \ref{sec::Headrick}, we 
introduce causality arguments which provide constraints in the covariant 
HRT-like prescription. In search of constraints for the static RT-like 
prescription, in section \ref{sec::GBconstraints} we turn to an alternative 
method of locating extremal surfaces for entanglement entropy in static 
scenarios, first proposed by Lewkowycz and Maldacena \cite{Lewkowycz:2013nqa}. 
In section \ref{sec::conclusion} we summarise our results, and discuss their 
implications for functional prescriptions of entanglement entropy. Several 
lengthy details are relegated to the appendices: In appendix \ref{sec::notation} 
we explain the notation used throughout the paper, and in appendix 
\ref{appendix::EOM} the full form of the equation of motion studied in section 
\ref{sec::NMGtotal} is given. In appendix \ref{appendix::BH} we briefly
discuss extremal surfaces for the rotating BTZ and Lifshitz black holes of NMG.

While in the final stages of this project, we learned of a similar study 
\cite{SINHA} where it is investigated under which conditions there might be a 
clash between holographic prescriptions for computing relative entropy and the 
manifest positivity of this quantity in field theory terms.

\section{Functional prescriptions of holographic entanglement entropy}
\label{sec::2}

Let us introduce the covariant and static functional prescriptions for the 
holographic calculation of entanglement entropy in Einstein-Hilbert gravity and 
general higher curvature theories. We utilise functional prescriptions for 
entanglement entropy throughout this paper, and only in section 
\ref{sec::GBconstraints} is an alternative prescription discussed, which is 
based on extending the replica trick into the bulk. We refer to this alternative 
method as the \emph{conical boundary condition method}.

\begin{figure}[htbp]                                 
\begin{center}
\includegraphics[width=0.4\linewidth]{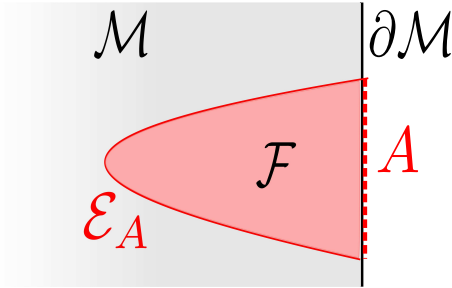}
\end{center}
\caption[]{Setup for the calculation of the holographic entanglement entropy 
corresponding to the boundary region $A$. The time direction of $\mM$ (and 
$\partial\mM$) is suppressed in this picture. 
}
\label{fig::RT}
\end{figure}

For Einstein-Hilbert gravity, the Ryu-Takayanagi (RT) prescription provides a 
means of holographically calculating entanglement  entropy in the static case: 
When the bulk spacetime $\mM$ with asymptotic boundary $\partial\mM$ is static, 
there exists a timelike Killing vector field that induces a foliation of both 
$\mM$ and $\partial\mM$ into spacelike surfaces. For a CFT region $A$ on such a 
spacelike slice of the boundary $\partial\mM$, the entanglement entropy 
$\mS_{A}$ associated with $A$ is given by the area of the bulk minimal-area 
surface $\mE_{A}$ located on the same spacelike slice in the bulk via
\begin{align}
 \mS_{A}=\frac{\text{Area}\left(\mE_{A}\right)}{4G_N},
 \label{RT}
\end{align}
where $G_N$ is the gravitational constant. In this formula, $\mE_A$ is anchored to 
$\partial \mM$, such that the 
intersection of $\mE_{A}$ with the boundary $\partial \mM$ equals the boundary 
$\partial A$ of the CFT-region $A$ (figure \ref{fig::RT}). The bulk minimal 
surface is also required to satisfy the \emph{homology condition}: 
there must exist a 
hypersurface $\mF$ in $\mM$ such that the boundary of $\mF$ is the union of 
$\mE_{A}$ and $A$ \cite{Fursaev:2006ih}. $\mE_{A}$ is then said to be 
homologous to $A$,\footnote{Note that in the static RT prescription $\mF$ is by 
construction located in the constant-time slice of $\mM$.} and we will 
refer to such surfaces as \textit{(holographic) entangling surfaces}.

The Hubeny-Rangamani-Takayanagi (HRT) prescription extends RT to arbitrary 
time-dependent states, thus providing a covariant prescription for holographic 
entanglement entropy. More concretely, for a given spacelike CFT region $A$ on 
the boundary $\partial \mM$ of the asymptotically AdS spacetime, one searches 
for bulk surfaces $\mE_{A}$ anchored to $\partial \mM$ which 
\emph{extremise} the area functional, picking the surface that gives the 
smallest entropy:
\begin{align}
 \mS_{A}= \text{min}_{X}\: 
\frac{\text{Area}\left(\mE_{A}\right)}{4G_N} \qquad X = \left\{ \mE_{A} \; : \; 
\partial \mE_{A} \equiv 
 \mE_{A} \cap \partial \mM = \partial A \right\} .
 \label{HRT}
\end{align}

It was recently noted in \cite{Hubeny:2013gta} that in order to obtain 
consistent results in certain bulk spacetimes, for example those with multiple 
bifurcation surfaces like the Reissner-Nordstr\"{o}m black hole, the homology 
condition has to be either refined or appended with an additional causality 
condition. The authors of \cite{Hubeny:2013gta} imposed the additional 
constraint that the hypersurface $\mF$ is spacelike, as was already done by 
Wall in \cite{Wall:2012uf}. This seems to follow naturally from the RT approach, 
where $\mF$ is spacelike by construction. 

Headrick, Hubeny, Lawrence and Rangamani in contrast obtain a causality 
constraint on the HRT prescription in \cite{HEADRICK,HUBENY}. 
Primarily, this causality constraint demands that there should be no causal 
contact possible between $\mE_A$ and $A$.\footnote{It 
would be an interesting question to investigate when the causality constraint 
and the spacelike constraint on $\mF$ are equivalent. For example, it is 
conceivable that in a spacetime where closed timelike curves are created by 
identifying two spacelike slices it is impossible to satisfy the causality 
condition, while the demand that $\mF$ should be spacelike could still be 
easily met. The relation between these two possible conditions will be 
investigated in \cite{HUBENY}. However we are not going to deal with causally 
pathological examples in this work, and we assume in the discussion of section 
\ref{sec::Headrick} that both conditions could be used equivalently.} This 
argument will be explained in more detail in section \ref{sec::Headrick}, and 
plays an instrumental role in the findings of this paper as it enables us to 
consistently rule out the unphysical additional surfaces in a Lorentzian 
setting. The fact that we can use this argument to exclude these surfaces is 
one 
of the main results of this work.

We will henceforth adopt the convention that the HRT prescription is based on 
the variation of the area functional using a general homology condition, as 
originally proposed in \cite{Hubeny:2007xt}. Whenever we make use of a refined 
homology condition or additional causality conditions, such as in section 
\ref{sec::Headrick}, we will explicitly mention this.

As mentioned in the introduction, for the holographic calculation of 
entanglement entropy in CFTs dual to higher curvature gravity theories, 
modified functionals have been proposed in
\cite{Fursaev:2013fta,Miao:2013nfa,Dong:2013qoa,Camps:2013zua}.  In particular, 
for a general four-derivative theory of gravity with (Lorentzian) Lagrangian
\begin{align}
S=\frac{1}{16\pi G_N}\int \! d^{d+1}x\,\sqrt{-g}\left[R+2\Lambda
+aR^2+bR_{\mu\nu}R^{\mu\nu}+cR_{\mu\nu\alpha\beta}R^{\mu\nu\alpha\beta}\right], 
\label{HCaction}
\end{align}

the following functional was first derived in \cite{Fursaev:2013fta}:
\begin{align}
\mS_{EE}=\frac{1}{4G_N}
\int_{\Sigma} \! d^{d-1}y \sqrt{\gamma} \,
\left[1+2aR+b\left(R_{\|}-\frac{1}{2}
k^2\right)+2c\left(R_{\| \|}-\text{Tr}(k)^2\right)\right ],
\label{generalfunctional}
\end{align}

where $\Sigma$ is a spacelike co-dimension two hypersurface, with induced 
metric $\gamma_{ij}$ and extrinsic curvature terms $k$ and $\text{Tr}(k)^2$. 
Further explanation for the notation used in this formula can be found in 
appendix \ref{sec::notation}. 

In using the entropy functionals \eqref{generalfunctional} to calculate 
entanglement entropy of a given CFT region $A$ holographically, a crucial step 
is to locate the particular co-dimension two surface anchored to $\partial A$ in 
the bulk upon which the functional should be evaluated.  Since for any given $A$ 
there is an infinite number of possible surfaces obeying the homology 
condition, a priori it is not clear which of these should be chosen. In 
principle it should be possible to determine the surface by solving all the 
equations  of motion of the higher curvature theory with conical boundary 
conditions \cite{Lewkowycz:2013nqa,Dong:2013qoa}, and this is demonstrated for 
Gauss-Bonnet gravity in section \ref{sec::GBconstraints}. However for a general 
higher curvature theory, in practice this calculation is very involved. As noted 
in \cite{Dong:2013qoa}, it would be advantageous to instead be able to determine 
the surface by directly extremising the functional. This was argued to be 
true for theories of the form \eqref{HCaction} by comparing the equations of 
motion derived from extremising the functional with certain equations derived 
from the conical boundary condition method in \cite{Dong:2013qoa}. In this 
paper we find that this method of locating the correct co-dimension two surfaces 
of the functional \eqref{generalfunctional}, for consistency and physicality, 
needs to be supplemented by constraints additional to the homology constraint on 
the extremal surfaces. Our conclusions are drawn by studying the nature of the 
extremal surfaces of the entropy functional \eqref{generalfunctional} for 
(partially) static backgrounds in New Massive gravity 
\cite{Bergshoeff:2009hq,Bergshoeff:2009aq} and Gauss-Bonnet gravity 
\cite{Lovelock:1971yv,Padmanabhan:2013xyr}.

Let us introduce some definitions employed throughout this paper, which 
translate the RT and HRT prescriptions of Einstein-Hilbert gravity to general 
higher curvature theories. When extremising the entropy functional 
\eqref{generalfunctional} on an equal time slice of a (partially) static 
spacetime, we refer to this as the \textit{RT-like prescription}. In the 
\emph{HRT-like prescription}, we need not restrict extremising the functional to 
equal time slices of the spacetime, as they may not be uniquely defined at all 
(e.g. when the extremal curves lie inside a black hole horizon as in 
section \ref{sec::closedsurfaces}). Whenever using causality arguments that 
intrinsically depend on a covariant setting this will be refered to as an 
\textit{HRT-like prescription amended with additional conditions}.

In Einstein-Hilbert gravity, for a (partially) static setting the RT 
prescription and the HRT prescription with additional conditions are believed to 
be equivalent. However in the higher curvature case, our findings show that when 
taking causality conditions into account, the RT- and HRT-like prescriptions 
do not agree at least in NMG and Gauss-Bonnet gravity and it is the HRT-like 
prescription amended with a causality constraint that gives the physically 
expected result, see section \ref{sec::Headrick}. We find these 
inconsistencies by investigating closed extremal surfaces of the entropy 
functionals in black hole and AdS backgrounds, whose significance we 
explain in the next 
section.

\section{Significance of closed extremal surfaces}
\label{sec::closedcurves}

\begin{figure}[htbp]                                 
\begin{center}
\includegraphics[width=0.69\linewidth]{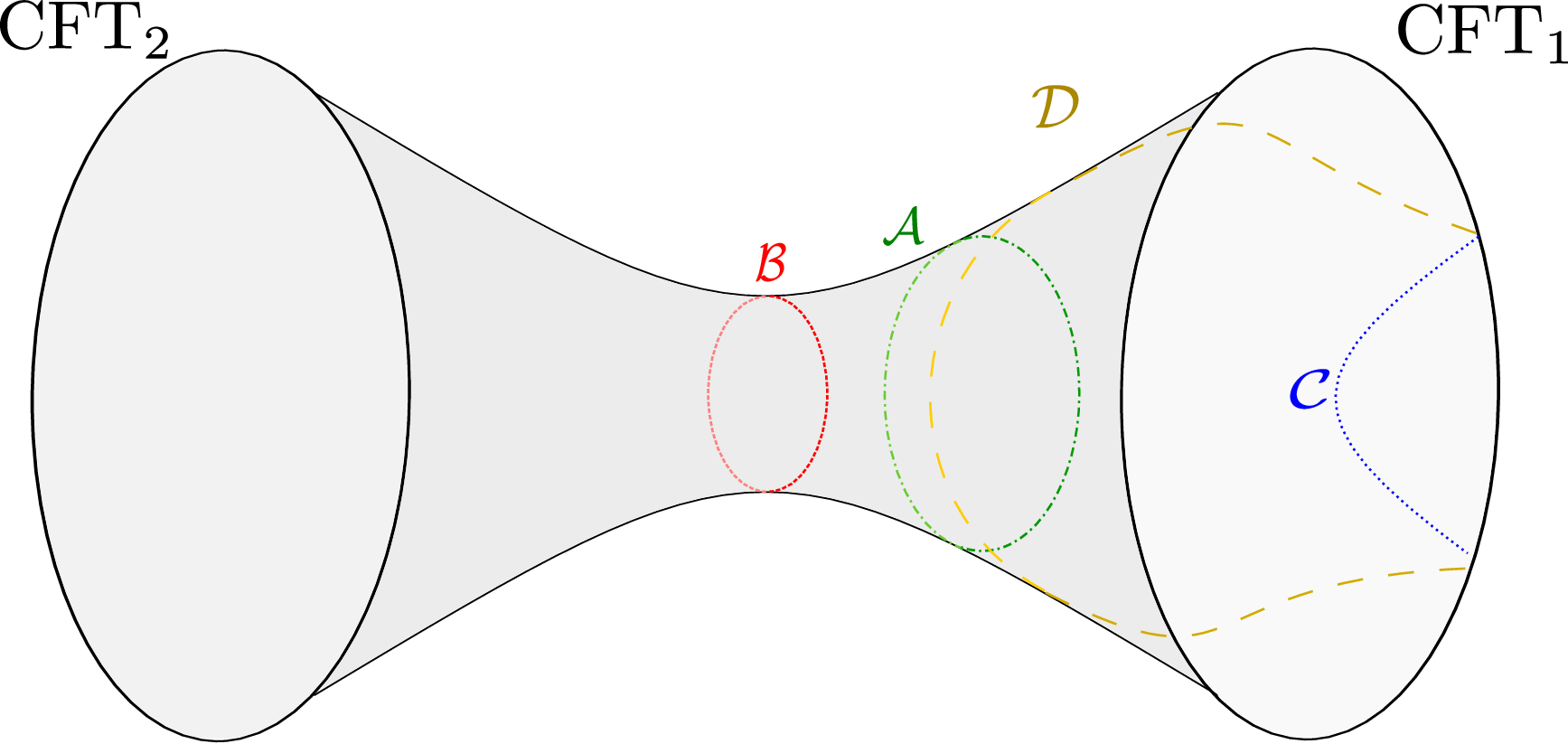}
\includegraphics[width=0.3\linewidth]{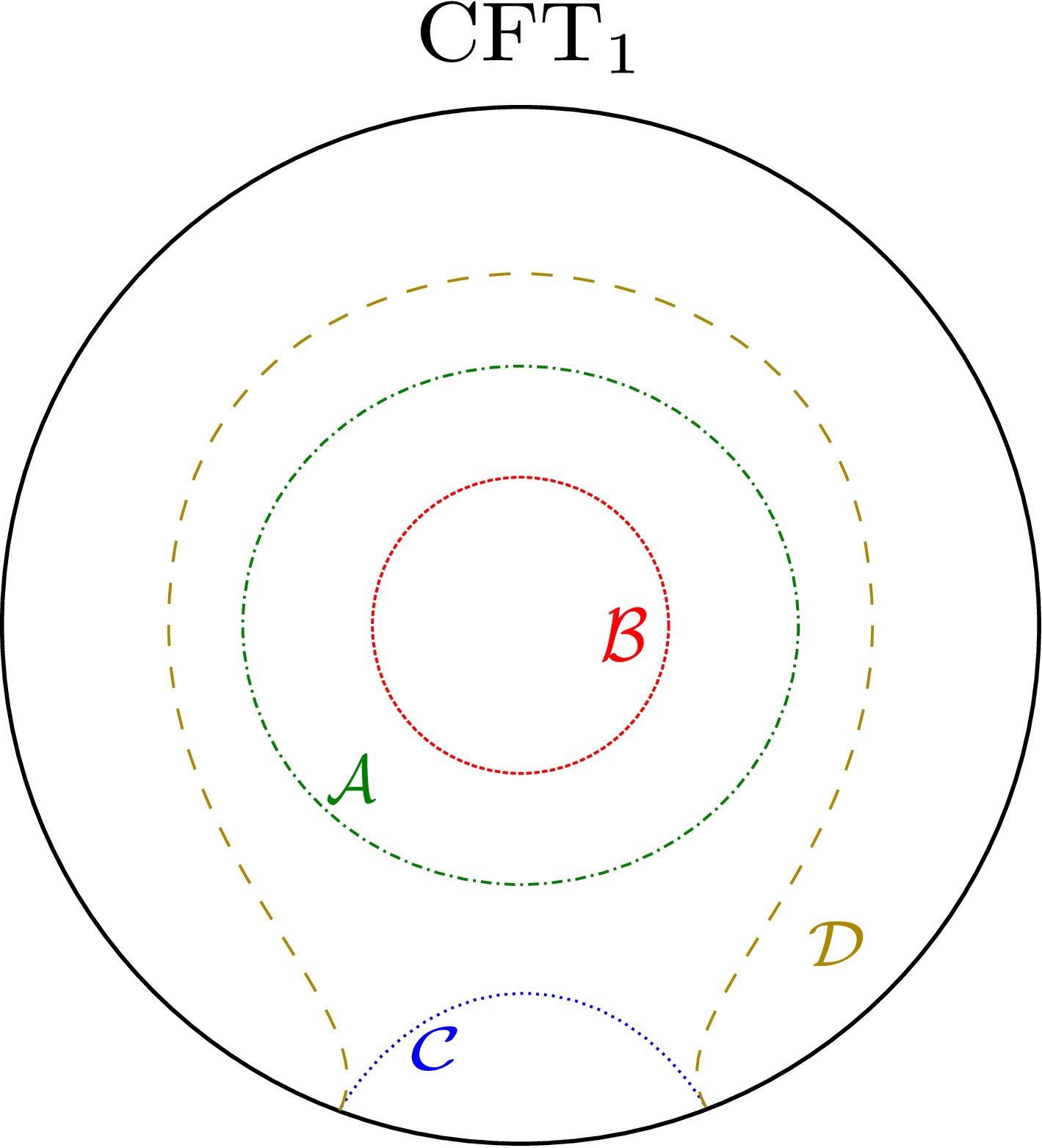}
\end{center}
\caption[]{On the left side, we sketch a spacelike slice of a black hole 
spacetime with $t=$const., also known as an Einstein-Rosen bridge. In a 
conformal diagram such as figure \ref{fig::conf}, this corresponds to a 
straight line through the center connecting the two asymptotic boundaries. On 
the right hand side, we sketch the (compact) boundary with $\text{CFT}_1$ and 
the interior of the bulk between the boundary and the bifurcation surface $\mB$. 
This surface $\mB$ is also an extremal surface. $\mC$ and $\mD$ sketch typical 
extremal surfaces anchored to the boundary, as found in 
\cite{Hubeny:2012wa,Hubeny:2013gta}, while $\mA$ depicts the possibility of an 
additional closed extremal surface that might appear in higher curvature 
theories.
}
\label{fig::2}
\end{figure}

When calculating the entanglement entropy of a (small) subsystem of the whole 
boundary using the RT and HRT prescriptions, one needs to search for 
surfaces extremising the entropy functional that are anchored at the asymptotic 
boundary. But there may also exist extremal surfaces which are not 
anchored to the boundary, and instead form closed surfaces in the bulk. 
This is illustrated for the example of a black hole background in 
\ref{fig::2}.

In \cite{Headrick:2007km,Hubeny:2013gta}, for stationary black 
hole backgrounds in Einstein-Hilbert gravity it was shown that these closed 
extremal surfaces are very important: They are homologous to the full boundary 
and thus determine the total entropy of the dual CFT. These surfaces play a 
role when larger regions of the boundary are considered \cite{Hubeny:2013gta}.

In Einstein-Hilbert gravity, closed extremal surfaces of the area functional 
only exist in black hole backgrounds. They are always the bifurcation 
surface(s) of the black hole, at least assuming the weak energy condition: 
$T_{\mu\nu}V^{\mu}V^{\nu}\geq0$ for any timelike vector $V^{\mu}$ (see 
\cite{Headrick:2013zda,HEADRICK}). The RT prescription, being restricted to 
equal time slices, will then naturally equate CFT entropy with the area of the 
outer (if an inner one exists at all) bifurcation surface, as expected from 
standard AdS/CFT lore \cite{Witten:1998zw}. In order to reproduce this result in 
the HRT prescription, as mentioned in the previous section, one has to impose 
additional constraints such as the causality constraint \cite{Hubeny:2013gta}. 
The CFT entropy is then correctly determined by the (outer) bifurcation surface 
$\mB$, and is then equal to the usual Bekenstein-Hawking black hole entropy
\begin{align}
 \mS_{BH}=\frac{\text{Area}(\mB)}{4G_N}.
\end{align}

\begin{figure}[htbp]                                 
\begin{center}            
\includegraphics[width=0.4\linewidth]{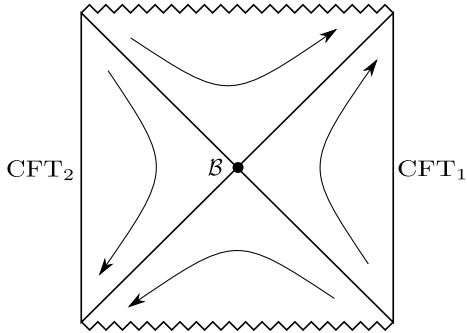} 
\end{center}
\caption[]{Conformal diagram of a static asymptotically AdS black hole. The 
bifurcation surface is denoted by $\mathcal{B}$, the singularities are drawn as 
zig-zag lines, the event horizons are the black diagonals and the timelike Killing 
vector field $\xi$ is denoted by the arrows.
}
\label{fig::conf}
\end{figure}

In higher curvature theories, we may argue by symmetry that black hole 
bifurcation surfaces will also be saddle points of any functional of the form 
\eqref{generalfunctional}: The bifurcation surface is defined by the vanishing 
of a Killing vector field $\xi$. By symmetry the functional, when evaluated on 
certain curves, has to be invariant under flows generated by this Killing 
field. Hence the bifurcation surface will define a saddle point of this 
functional in spacetimes with the typical global structure of a black hole, see 
figure \ref{fig::conf}. It is an important consistency check that when 
evaluated on the (outer) bifurcation surface of a black hole, the functional 
\eqref{generalfunctional} reproduces Wald's formula 
\cite{Wald:1993nt,Jacobson:1993vj,Iyer:1994ys} for black hole 
entropy in higher curvature theories.

However, due to the complexity of the functional \eqref{generalfunctional}  
compared to the area functional of Einstein-Hilbert gravity, it is possible for 
other closed extremal surfaces to exist, even in non-black hole 
backgrounds. For a black hole geometry, this is sketched in figure 
\ref{fig::2}. A priori it is not clear in this case whether the RT-like 
and HRT-like prescriptions will both correctly identify the (outer) bifurcation 
surface as the closed extremal curve associated to total CFT entropy. This is 
addressed in section \ref{sec::Headrick}, after investigating the existence of 
such surfaces in NMG and 4+1-dimensional Gauss-Bonnet gravity in the following 
sections.

\section{Extremal curves in New Massive gravity}
\label{sec::NMGtotal}

\subsection{NMG and proposed entropy functional}
\label{sec::NMG} 

The action for New Massive gravity (NMG) with cosmological parameter $\lambda$ 
in its Lorentzian form is given by
\cite{Bergshoeff:2009hq,Bergshoeff:2009aq}
\begin{align}
S=\frac{1}{16\pi G_N}\int \! d^3x\sqrt{-g}\,\left[\sigma R-2\lambda m^2
+\frac{1}{m^2}\left(R_{\mu\nu}R^{\mu\nu}-\frac{3}{8}R^2\right)\right],
\label{NMG}
\end{align}
where the mass parameter $m^2$ can take any sign, and $\sigma=\pm1$
 is the sign of the  Einstein-Hilbert term. This theory has been 
investigated in a holographic context for example in 
\cite{Liu:2009bk,Grumiller:2009sn,Liu:2009kc,dS:2010fh,Sinha:2010ai,
Chen:2013aza,Alishahiha:2013zta,Chen:2014kja}. For later convenience, some 
important properties of this theory are detailed in the following 
\cite{Bergshoeff:2009aq}:

For  $\lambda \ge -1$ the equations of motion from the above action admit 
maximally symmetric vacua as solutions. In particular, NMG admits AdS$_{3}$ 
vacua with the  AdS radius $\ell$ (and $\Lambda=-1/\ell^{2}$) determined by the 
real solutions of the equation 
\begin{align}
\frac{1}{\ell^2}=2m^2\left(\sigma\pm\sqrt{1+\lambda}\right).
\label{AdS}
\end{align}
When linearising around a maximally symmetric background with curvature 
$\Lambda$, negative energy gravitons (ghosts) are avoided when 
\begin{align}
 m^2(\Lambda-2m^2\sigma)>0,
 \label{ghosts}
\end{align}
while the Breitenlohner-Freedman (BF) bound reads
\begin{align}
 -2m^2\sigma\geq\Lambda.
 \label{BF}
\end{align}
By using the Brown-Henneaux reasoning \cite{Brown:1986nw} on higher 
curvature theories, the central charges of the dual CFT of NMG have 
been determined to be \cite{Kraus:2005vz,Bergshoeff:2009aq}
\begin{align}
 c=\frac{3\ell}{2 G_N}\left(\sigma+\frac{1}{2m^2\ell^2}\right),
 \label{cc}
\end{align}
implying a clash between unitarity and positive energy in the bulk, and positive 
central charge of the boundary CFT: the ghost-free condition \eqref{ghosts} and 
the condition $c\geq0$ are mutually exclusive. Amongst other solutions, NMG 
gravity admits the BTZ black hole \cite{Banados:1992gq,Banados:1992wn}, whose 
entropy is proportional to the above central charge 
\cite{Kraus:2005vz,Bergshoeff:2009aq}: 
\begin{align}
\mS_{BTZ}=\frac{2\pi r_+}{4 G_N}\left(\sigma+\frac{1}{2m^2\ell^2}\right).
\label{S_BTZ}
\end{align}
 $r_+$ is the radius of the (outer) event horizon. Positivity of the BTZ black 
hole entropy hence demands the positivity of the central charge.

For NMG, the functional \eqref{generalfunctional} for the holographic 
calculation of entanglement entropy reduces to 
\cite{Bhattacharyya:2013gra}
\begin{align}
\mS_{EE}=\frac{1}{4 G_N}\int \! d\tau\sqrt{g_{\tau\tau}}\,\left[\sigma
+\frac{1}{m^2}\left(R_{\|}-\frac{1}{2}k^2-\frac{3}{4}
R\right)\right ],
\label{SEE}
\end{align}

whose evaluation on a particular extremal surface (or curve, since we are in 
2+1 dimensions) is proposed to compute entanglement entropy in the dual boundary 
theory. The integral is performed along a curve parametrised by $\tau$, with the 
induced metric 
$\sqrt{g_{\tau\tau}}=\sqrt{g_{\mu\nu}\frac{dx^{\mu}}{d\tau}\frac{dx^{\nu}}{d\tau
} } $. In Einstein-Hilbert gravity, the entropy functional in 2+1 dimensions 
just computes the length of a path, and its extremisation produces the geodesic 
equations of motion. However in NMG, the evaluation of functional \eqref{SEE} no 
longer has the interpretation of length, due to the presence of additional 
curvature terms.

In the following we investigate the nature of the extremal curves corresponding 
to \eqref{SEE} in (partially) static backgrounds by deriving and solving 
the equations of motion. In the light of section \ref{sec::closedcurves} we pay 
particular attention to closed extremal curves.

\subsection{Equations of motion of NMG entropy functional}
\label{sec::EOM}

As we mentioned in section \ref{sec::2}, in using the functionals \eqref{generalfunctional} 
to calculate entanglement entropy holographically, first the correct co-dimension two 
surface upon which the functional is evaluated needs to be found.
 It is hoped that one way of locating these surfaces is by directly 
extremising the functionals \cite{Dong:2013qoa}, and in light of this we 
investigate the possible curves which extremise \eqref{SEE}, considering global 
AdS$_{3}$ and non-rotating BTZ black hole backgrounds.
In 2+1 dimensions, these metrics take the form:
\begin{equation}
ds^{2} = -\left(-M+\frac{r^{2}}{\ell^{2}}\right)dt^{2}+\left(-M+\frac{r^{2}}{\ell^{2}}\right)^{-1}dr^{2}+r^{2}d\phi^{2} 
\label{nonrotBTZ}
\end{equation}
in Schwarzschild coordinates. The global AdS$_{3}$ metric is obtained by setting 
the BTZ black hole mass $M=-1$. We obtain the curves which extremise 
\eqref{SEE} as follows: 

By considering curves that lie in a constant time slice of 
$t=0$,\footnote{This can be assumed as the vector $\partial_{t}$ is a Killing 
vector in all the spacetimes we will work with.} we may choose the 
parameterisation $r=f(\phi)$, i.e. the progression of the curve into the bulk 
spacetime is given as a function of the (boundary) coordinate $\phi$. This can 
be inserted into \eqref{SEE}, giving
\begin{align}
\mS_{EE} =\frac{1}{4 G_N} \int \! d\phi \,
\; \sigma\left(f\left(\phi \right)^2+\frac{f'\left(\phi \right)^2}{\frac{f\left(\phi \right)^2}{\ell ^2}-M}\right)^{\frac{1}{2}}
\left(1 +\frac{1-\ell^2k^2\left(\phi\right)}{2 \sigma m^2 \ell 
^2}\right)
 \label{EOM_NMG}
\end{align}
 with the extrinsic curvature term
\begin{align}
k^2\left(\phi\right)&=-\frac{1}{\ell ^2 \left(M \ell ^2 
f\left(\phi \right)^2-f\left(\phi \right)^4-\ell ^2 f'\left(\phi 
\right)^2\right)^3}
\Big{(}-2 M \ell ^2 f\left(\phi \right)^4+f\left(\phi \right)^6
\nonumber
\\
&-2 M \ell ^4 
f'\left(\phi \right)^2+f\left(\phi \right)^2 
\left(M^2 \ell ^4+3 \ell ^2 f'\left(\phi \right)^2\right)+M \ell ^4 f\left(\phi 
\right) 
f''\left(\phi \right)-\ell ^2 f\left(\phi \right)^3 f''\left(\phi 
\right)\Big{)}^2.
\nonumber
\end{align}
The corresponding fourth order Euler-Lagrange equation of motion for $f(\phi)$ 
is quite involved, and we relegate the explicit expression to equation 
\eqref{fourthorder} of appendix 
\ref{appendix::EOM}. 

In following we discuss this equation and its possible solutions, which was in 
part already done in \cite{Alishahiha:2013zta}. We begin with solutions anchored 
to the boundary.

\subsection{Curves anchored at the boundary}
\label{sec::anchored}

The fourth order nature of equation \eqref{fourthorder} arises from the 
presence of the extrinsic curvature term $k^2$ in \eqref{SEE}, and therefore to 
find a unique solution when solving this equation we have to specify initial 
conditions up to third order in derivatives, i.e. values 
$f(\phi_0),f'(\phi_0),f''(\phi_0),f'''(\phi_0)$. However it is first 
interesting to note that the geodesics of background \eqref{nonrotBTZ}, 
which are used to calculate entanglement entropy holographically in 
Einstein-Hilbert gravity, solve \eqref{fourthorder} independently of the NMG 
$m^2$ parameter\footnote{This is likely a consequence of the fact 
that an interval is merely the 1-ball. In \cite{Casini:2011kv} an argument was 
presented that when the boundary region is a ball, the corresponding 
holographic 
entangling surface in AdS space can be found by an argument using topological 
black holes, and is independent of the bulk gravity theory.}. For global 
AdS$_{3}$ this was already noted in 
\cite{Alishahiha:2013zta}. Setting $\ell = 
1$, these 
geodesics are \cite{Hubeny:2013gta,Hubeny:2012wa}

\begin{equation}
f(\phi)=\frac{\cos(\phi_0)}{\sqrt{\cos(\phi)^2-\cos(\phi_0)^2}}
\label{geoAdS}
\end{equation}
for global AdS$_{3}$, and 
\begin{align}
f(\phi)=r_{+}\left(1-\frac{\cosh^2(r_{+}\phi)}{\cosh^2(r_{+}\phi_0)}\right)^{
-\frac{1}{2} } 
\label{geoBTZ}
\end{align}
for a non-rotating BTZ background, with event horizon radius $r_{+} = \sqrt{M}$. 

\begin{figure}[htb]                                 
\begin{center}                                  
\includegraphics[width=0.45\linewidth]{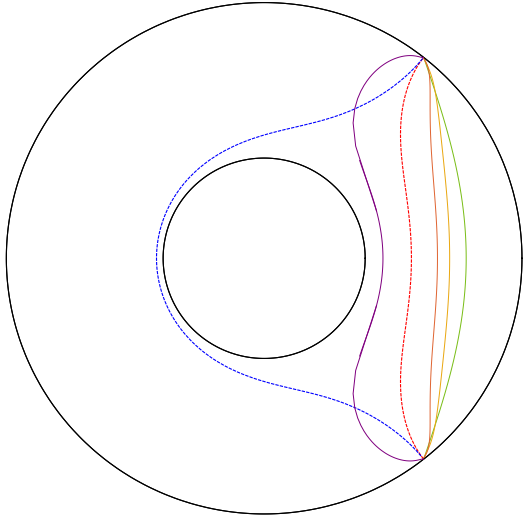}
\end{center}
\caption[]{Diagram of a spacelike slice of the spacetime \eqref{nonrotBTZ} for
the choice of parameters $\sigma=\ell=1, M=1/2, m^2=1/4$. The
outer (black) circle is asymptotic infinity, mapped to a finite radius via 
$r\rightarrow\arctan(r)$. The inner (black) circle is the bifurcation surface, 
the additional closed extremal curve in the bulk is not
shown. The two dashed curves are geodesics \eqref{geoBTZ} and hence solutions
of \eqref{fourthorder} that are independent of $m^2$, see text. The other
curves depicted are some of the other non-geodesic solutions to
\eqref{fourthorder}, anchored to the same boundary points.
}
\label{fig::multiplecurves}
\end{figure}

Other solutions can be found numerically, solving for curves with a turning 
point $f'(\phi_0) = 0$ at some initial radius $f(\phi_0)$ in the bulk. This 
allows us by symmetry to set $f'''(\phi_0) = 0$, leaving the freedom to specify 
$f''(\phi_0)$. In global AdS$_{3}$, we find non-geodesic curves anchored to the 
boundary when the collective NMG parameter $\sigma \ell^{2} m^{2}$ is 
negative. In fact, for a given finite interval $A$ on the boundary, for this 
parameter range there seems to exist an infinite number of different curves 
attached to $\partial A$. For the non-rotating BTZ black hole background, we 
again appear to find an infinite number of curves anchored to a given boundary 
region 
for $\sigma \ell^{2} m^{2} > 0$, figure \eqref{fig::multiplecurves}.

The freedom to specify an initial condition on $f''(\phi_0)$ in the above 
appears to be behind the emergence of the multiplicity of extremal curves 
associated to a given boundary interval. Notably, there also appears to 
be an infinite subset of these curves which satisfy the homology 
constraint. Given the higher derivative nature of the equations 
\eqref{fourthorder}, it is likely that a sufficient set of boundary conditions 
would select the appropriate curve for entanglement entropy. We leave the 
investigation of correct boundary conditions on these holographic entangling 
curves for future research, as it is beyond the scope of our paper. In any case, 
in section \ref{sec::Headrick} we will present another consistency argument that 
also constrains this infinite set of curves in such a way that only the geodesic 
solutions remain.

In the next section we will consider the possible closed extremal curves, 
which we are to study analytically. The corresponding value of entropy can then 
be easily calculated as no boundary terms are required.

\subsection{Closed extremal curves in NMG}
 \label{sec::closedsurfaces}

For the study of closed extremal curves, to begin we restrict our 
attention to black hole backgrounds. In particular, for this section we focus on 
the non-rotating BTZ black hole while in appendix \ref{appendix::BH} we comment 
on the rotating case, and Lifshitz black holes. In section 
\ref{sec::closedcurves} the significance of closed extremal curves of 
holographic entropy functionals in black hole backgrounds was explained.

While the Schwarzschild coordinates \eqref{nonrotBTZ} do not cover the BTZ black 
hole event horizon, following the discussion of section 
\ref{sec::closedcurves} the BTZ black hole bifurcation surface $r_{+} = 
\ell \sqrt{M}$ is a closed extremal surface of entropy functional \eqref{SEE}. 
The corresponding value of the entropy functional is equal to the entropy of 
the black hole \eqref{S_BTZ}:
\begin{align}
\mS_{EE+}=\mS_{BTZ}=\frac{2\pi\ell\sqrt{M}}{4 G_N}
\left(\sigma+\frac{1}{2\ell^2m^2}\right).
\end{align}

\begin{figure}[htbp]                                 
\begin{center}            
\includegraphics[width=1\linewidth]{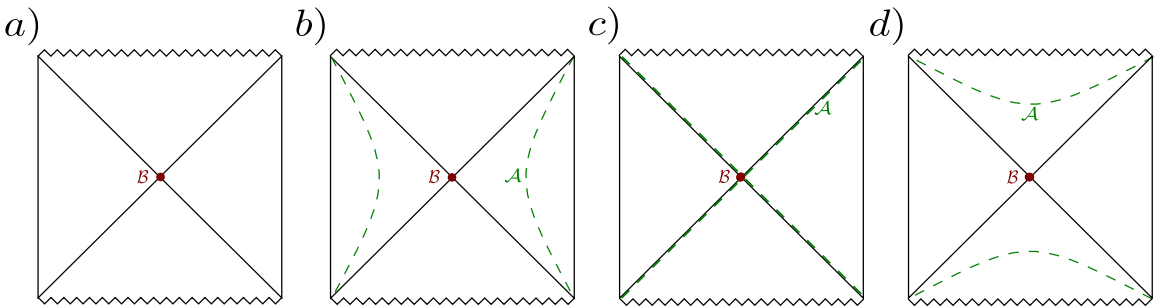}
\end{center}
\caption[]{$a)$: Conformal diagram of the BTZ black hole \eqref{nonrotBTZ} 
without additional extremal surfaces as also in figure \ref{fig::conf}. The 
bifurcation surface is denoted by $\mathcal{B}$, the singularities are drawn as 
zig-zag lines and the event horizons are the black diagonals. The appearance of 
additional closed extremal surfaces [$b)-d)$] is not entirely a feature of 
the 
geometry (as geodesics would be), but also depends on the parameters of the 
gravitational theory. Additional extremal surfaces ($\mathcal{A}$) can be 
outside of the event horizon [$b)$], inside of it [$d)$] or in special cases 
they can coincide with slices of the event horizon [$c)$].
}
\label{fig::1}
\end{figure}

\begin{figure}[htbp]                                 
\begin{center}                                  
\includegraphics[width=0.8\linewidth]{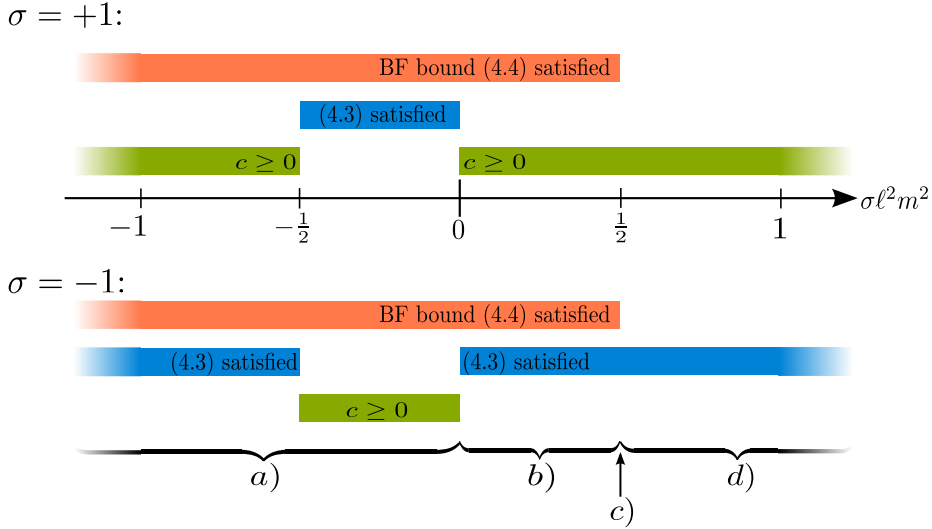}  
\end{center}
\caption[]{An overview over the parameter-space of NMG: For the two possible 
choices $\sigma=+1$ and $\sigma=-1$, it is depicted for which values of 
$\sigma\ell^2 m^2$ the inequalities \eqref{ghosts} and \eqref{BF} are satisfied 
and \eqref{cc} yields positive central charges. Comparing \eqref{additional} 
and $r_{+}=\ell\sqrt{M}$, we can depict for which value of $\sigma\ell^2 m^2$ 
what type of additional extremal surface (see $a)-d)$ in figure \ref{fig::1}) 
appears. 
}
\label{fig::3}
\end{figure}

To find closed curve solutions to the equation of motion \eqref{fourthorder}, 
we set $f'=f''=f'''=f''''=0$. For the same range that the 
multiplicity of extremal curves anchored to the boundary 
appears in the non-rotating BTZ background $\left(\sigma\ell^2 m^2 > 
0\right)$, we find an additional closed extremal curve 
located at radial distance\footnote{Performing the same calculation in Kruskal 
coordinates, one can show that this also holds when the additional  closed 
extremal curves are inside the event horizon. }
\begin{align}
 r_{a}=\sqrt{\frac{M}{2\sigma m^2}}
 \label{additional}
\end{align}
from the centre. Notice that $r_a\geq r_{+}$ is equivalent to the BF bound 
\eqref{BF}. Our results for these additional closed bulk extremal curves are 
summarised in figures \ref{fig::1} and \ref{fig::3}. The value of the entropy 
functional corresponding to $r_{a}$ is
\begin{align}
\mS_{EEa}=\frac{2\pi\sigma}{4 G_N}
\sqrt{\frac{2M}{\sigma m^2}}.
\label{additionalEntropy}
\end{align}

Comparing this value with the entropy of the BTZ black hole $\mS_{BTZ}$, we find 
$|\mS_{BTZ}|\geq|\mS_{EEa}|$ with 
equality for $\sigma\ell^2 m^2=1/2$ where $r_a=r_+$, i.e. where the BF 
bound is saturated. For positive central charges (and hence positive BTZ 
entropy), the additional extremal surfaces will correspond to a lower entropy 
than the Wald entropy of the black hole. Hence in both RT- and HRT-like 
prescriptions, naively invoking only the homology constraint and that the 
entropy is minimised would lead one to incorrectly identify $\mS_{EEa}$ with the 
entropy of the dual CFT for certain choices of $\sigma\ell^2m^2$. We elaborate 
on how this changes in the HRT-like prescription when causality constraints are 
imposed in section \ref{sec::Headrick}.

The existence of additional closed extremal curves is not restricted to the 
non-rotating BTZ background: they also exist for the rotating BTZ and Lifshitz 
black hole backgrounds, see appendix \ref{appendix::BH}.

As a further simple example let us consider global AdS spacetime, which can 
be obtained by setting $M=-1$ in the BTZ metric \eqref{nonrotBTZ}. It follows 
from equations \eqref{additional} and \eqref{additionalEntropy} that we also 
get additional closed extremal curves in the bulk of AdS for 
$\sigma\ell^2m^2<0$, which is case $a)$ in figure \ref{fig::3}. The 
corresponding entropy will be positive for $\sigma=+1$ and negative for 
$\sigma=-1$. In the case of negative entropy, these additional curves hence 
correspond to an entropy lower than the value which is physically expected. As 
in the above, we would be led to incorrectly identify this with the CFT entropy. 
It can be seen from figure \ref{fig::3} that part of this parameter range (for 
example $\sigma=-1$ and $\sigma\ell^2m^2<-1/2$) is free of ghosts, but the 
central charge of the dual theory would be negative. We will return to the 
issue 
of closed extremal curves in an AdS bulk spacetime in section \ref{sec::GBG}. In 
section \ref{sec::Headrick} we will also explain how the causality argument can 
be employed to rule out these additional curves.

As NMG is afflicted with problems regarding ghosts in the bulk (violation 
of inequality \eqref{ghosts}), one might ascribe the additional closed curves 
we find in the theory to instabilities of the background. However, the parameter 
ranges in which the additional curves exist and that in which 
ghosts appear are not in one-to-one correspondence (see figure \ref{fig::3}).
For example, global AdS space can have additional closed curves without 
ghosts, while for the BTZ black hole at $\sigma=+1,\  \sigma\ell^2m^2=-1$ the 
theory exhibits ghosts and does not show additional extremal curves. There is 
therefore no obvious and transparent connection between the appearance of 
ghosts and the additional curves.

In the following section, as another example 
we consider the existence of additional closed extremal surfaces in Gauss Bonnet 
gravity.

\section{Closed extremal surfaces in Gauss-Bonnet gravity}
\label{sec::GBG}

\subsection*{Gauss-Bonnet gravity}

Gauss-Bonnet gravity is a special case of Lovelock 
gravity \cite{Lovelock:1971yv} (see \cite{Padmanabhan:2013xyr} for a review) and 
has been extensively studied before in the holographic context, 
see for example \cite{Hung:2011xb,deBoer:2011wk} and
\cite{Camanho:2013pda,Edelstein:2013sna} for reviews. For 
simplicity we will restrict our discussion to five bulk dimensions, 
in which the (Lorentzian) action reads
\begin{align}
S=\frac{1}{16\pi G_N}\int \! d^5x\sqrt{-g}\,\left[R+\frac{12}{L^2}
+\lambda 
\frac{L^2}{2}\left(R_{\mu\nu\alpha\beta}R^{\mu\nu\alpha\beta}-4R_{\mu\nu}R^{
\mu\nu}+R^2\right)\right]
\label{GBaction}
\end{align}
adopting the conventions used for example in \cite{deBoer:2011wk}. The 
conjectured boundary theory is causal for 
\cite{Brigante:2007nu,Brigante:2008gz,Buchel:2009tt,Hofman:2009ug}\footnote{
Similar conditions on the value of $\lambda$ have been derived in 
\cite{SINHA} by demanding positivity of holographic relative entropy.}
\begin{align}
 -\frac{7}{36}\leq\lambda\leq\frac{9}{100},
\end{align}
and it is possible to choose an AdS vacuum such that the bulk theory is 
ghost free \cite{Boulware:1985wk} and the dual CFT is unitary (see e.g. 
\cite{Camanho:2013pda,Edelstein:2013sna}). The entropy functional 
\eqref{generalfunctional} takes the form
\begin{align}
\mS_{EE}=\frac{1}{4G_N}
\int_{\Sigma} \! d^3y \sqrt{\gamma}\, \left(1+\lambda L^2\mR\right),
\label{GBSEE}
\end{align}
where the Gauss-Codazzi equations have been used to absorb the extrinsic 
curvature terms into the intrinsic scalar curvature $\mR$ of $\Sigma$. 
This functional, known as Jacobson-Myers functional, has already been derived 
in \cite{Jacobson:1993xs} in the context of black hole entropy. It was later 
proposed for the holographic calculation of entanglement entropy in 
\cite{Fursaev:2013fta,Fursaev:2006ih,Hung:2011xb,deBoer:2011wk}.

\subsection*{Closed extremal surfaces}

At the level of the entropy functional it is possible to conclude that
 spherically symmetric spacetimes in Gauss-Bonnet gravity also admit additional 
closed extremal surfaces: Suppose we are working with a stationary spherically 
symmetric black hole background of Gauss-Bonnet gravity (see e.g 
\cite{Wheeler:1985nh,Wheeler:1985qd,Cai:2001dz,Camanho:2011rj}), given in 
Schwarzschild-like coordinates $t,r,\theta,\phi,\psi$. Similar to section 
\ref{sec::closedsurfaces}, we assume that the spacelike surface $\Sigma$ is 
adopted from the spherical symmetry of the spacetime, i.e. that $\Sigma$ is a 
3-sphere parametrized by $t=$const., $r=$const., with $\theta,\phi,\psi$ 
arbitrary. It then follows that 
$\int_{\Sigma}\! \sqrt{\gamma}\,=\text{Area}(\Sigma)=2\pi^2 r^3$ and 
$\mR=6/r^2$, such that the functional 
\eqref{GBSEE} takes the form
\begin{align}
 \mS_{EE}=\frac{\pi^2}{2G_N}\left(r^3+6\lambda L^2 r\right).
 \label{GBSEE2}
\end{align}

For negative $\lambda$ this has a minimum at finite 
$r=\sqrt{-2\lambda}L$, implying the existence of a closed extremal surface at 
this radius.\footnote{
It was recently shown 
\cite{Bhattacharyya:2013jma,Chen:2013qma,Bhattacharyya:2013gra,Pal:2013fha}, 
that the equations of motion originating from functionals that 
depend only on intrinsic curvature terms (such as \eqref{GBSEE}), take the 
general form $X^{ij}k^{(\alpha)}_{ij}=0$, where $k^{(\alpha)}_{ij}$ describes 
the extrinsic curvature projected onto $\Sigma$. For Gauss-Bonnet gravity, we 
have $X^{ij}=\frac{1}{2}\gamma^{ij}+\lambda 
L^2\left(\frac{1}{2}\gamma^{ij}\mR-\mR^{ij}\right)$. It is then easy to show 
that the additional closed bulk extremal surface at $r= \sqrt{-2 \lambda}L$ 
exactly solves these equations of motion with 
$X^{ij}=0$. See also section \ref{sec::GBconstraints} for a discussion 
of these equations in the context of the alternative conical boundary 
condition method.}\footnote{The case $\lambda<0$ is less studied, for in 
the context of string-theory one is restricted to $\lambda>0$ 
\cite{Boulware:1985wk}. } From a similar 
calculation as above, we see that for $\lambda>0$ additional extremal surfaces 
appear in hyperbolic backgrounds (where $\mR<0$) at a radius 
$r=\sqrt{2\lambda}L$. We now consider these additional extremal surfaces in a 
few simple backgrounds.

\subsection*{AdS and boson stars}

For spherically symmetric spacetimes such as global AdS or 
boson stars \cite{Hartmann:2013tca,Henderson:2014dwa},\footnote{Although the 
given sources only investigate boson stars for $\lambda > 0$, we were assured 
by Betti Hartmann and Yves Brihaye that similar solutions can also be found for 
$\lambda < 0$.} the additional surfaces 
are problematic. In these cases the additional surface 
would be competing with the empty surface, which corresponds 
to zero entropy. As the additional surface found above corresponds to 
negative entropy, naively following the prescription to take the surface 
extremising the functional with lowest entropy as the entangling surface would 
lead to erroneously prescribing a negative entropy to the CFT duals of AdS 
space and boson stars. Yet, as will be shown in section 
\ref{sec::Headrick} the 
causality argument presented will be sufficient to exclude these surfaces, so 
that the entropy of the CFT dual to an AdS or boson star spacetime is correctly 
identified as zero.

\subsection*{Black holes}

Note that the above argument for closed extremal surfaces is independent of the 
topology of the bulk spacetime, therefore in a stationary black hole 
spacetime, in addition to the surface found above, the 
black hole bifurcation surface would also be a closed extremal surface. 
For ghost free black hole backgrounds, the additional closed surface we 
find is not necessarily an issue due to the existence of lower bounds on the 
horizon radius which are larger than the radius of the additional surface. For the 
case of vacuum black holes \cite{Camanho:2011rj,Cai:2001dz} and charged 
hyperbolic black holes \cite{Torii:2005nh}, this bound on the event horizon 
exactly coincides with the radius of the 
additional surface. This means that these additional surfaces 
can never be outside of the event horizon, in the static region of the bulk 
spacetime. These black hole solutions can also have curvature 
singularities at finite radial coordinate. In the vacuum case 
\cite{Camanho:2011rj,Cai:2001dz}, this singularity is located at a larger 
value of the radial coordinate than the additional surface, the latter one hence 
does not appear in the spacetime at all. For the charged solutions 
\cite{Torii:2005nh} in contrast, the singularity can be located at a sufficiently 
low value of the radial coordinate such that the additional extremal surfaces appear 
at least inside of the outer event horizon, as in figure \ref{fig::1}$d)$. As 
the region inside the black hole is not static, this would be a 
problem in the HRT-like prescription.

\section{Closed bulk extremal surfaces and the causal influence argument}
\label{sec::Headrick}

In this section, we will present an argument based on causality that can 
be employed in Lorentzian settings to consistently eliminate the additional 
extremal surfaces found to be problematic throughout this paper.

\subsection*{The causal influence argument}

We first explain that the presence of the additional closed 
extremal surfaces found in this paper indicates a subtle difference 
between RT- and HRT-like 
prescriptions, when causality conditions are taken into account.\footnote{As 
explained before, following \cite{Dong:2013qoa}, in the RT- and HRT-like 
prescriptions we adopt the method of locating the entangling surface which 
involves extremising the entropy functionals.} More specifically, we will show 
that in HRT-like prescriptions, an argument based on causality can be used to 
argue that these surfaces cannot be used to calculate entanglement 
entropy.

In section \ref{sec::2} we explained how the area functional can be employed to 
calculate holographic entanglement entropy in Einstein-Hilbert gravity, with 
subtle differences between the RT and HRT prescriptions that could be 
resolved by imposing causality conditions on the latter, see also 
\cite{Hubeny:2013gta,HEADRICK}. Spacetimes for which such differences were 
found in \cite{Hubeny:2013gta} are the Reissner-Nordstr\"{o}m black hole and 
the bag-of-gold spacetime, where an eternal black hole is matched to a compact 
spacetime bubble along a static shell of matter 
\cite{Freivogel:2005qh,Marolf:2008tx}.

As was furthermore explained in section \ref{sec::2}, for higher curvature 
theories such as \eqref{HCaction}, functionals \eqref{generalfunctional} 
generalising the area functional have been proposed to calculate entanglement 
entropy holographically. As these functionals where derived in Euclidean 
settings in \cite{Fursaev:2013fta,Miao:2013nfa,Dong:2013qoa,Camps:2013zua}, a 
priori these functionals should be seen to be most reliable in an RT-like 
setting. However, as all the spacetime backgrounds that are considered in 
the present work are at least partially (i.e. not necessarily globally) 
static, one would naively expect RT- and HRT-like prescription in using these 
functionals to agree. Here we will now consider how the imposition of a 
causality argument due to Headrick, Hubeny, Lawrence and Rangamani \cite{HEADRICK,HUBENY}, which we refer 
to as the \emph{causal influence argument}, affects this expectation in our 
cases:

The principle behind the causal influence argument is to avoid causality 
paradoxes similar to the well known grandfather paradox. Suppose one wanted to 
calculate the entanglement entropy of a certain boundary region $A$ (at boundary 
time $t=0$) using a bulk co-dimension two surface $\mE_A$. If $\mE_A$ were to 
lie in the future of $A$, i.e. that there were to be a future pointing timelike 
curve from $A$ to (at least one point on) $\mE_A$,  the following paradox might 
arise: An observer living on (or near) the boundary might immediately after 
$t=0$ send some energy into the bulk in such a way that the geometry around the 
part of $\mE_A$ in the future of $A$ is changed. This would also affect the 
surface $\mE_A$ itself, hence potentially the associated entropy. This has the 
implication that it would be possible --- after the entanglement entropy was 
fixed from the point of view of the CFT --- to alter the result of the 
holographic entanglement entropy calculation in the bulk, leading to an 
obvious paradox. We therefore demand that there should be no timelike curve from 
$A$ to $\mE_A$, and by time inversion symmetry also not from $\mE_A$ to $A$. 
Similarly, the same holds for $\mE_A$ and $\bar{A}$, the complement of $A$. 
Hence, causality implies that $\mE_A$ should be required to lie in the 
\textit{causal shadows} of the boundary regions $A$ and $\bar{A}$, i.e. there 
should be no timelike curves connecting $\mE_A$ to one of the two 
regions\footnote{A note on nomenclature: The \textit{causal shadow} 
of a certain spacetime as originally defined by Headrick, Hubeny, Lawrence 
and Rangamani \cite{HEADRICK} is 
supposed to be the set of points in the bulk which are not causally connected to 
any point in the boundary, irrespective of any division of the boundary into 
subsystems $A$ and $\bar{A}$. Here we use this term also to define regions of 
the bulk which are not in causal contact with specific subregions of the 
boundary.}. For reasons that will soon become clear, we refer to this as the 
\textit{weak form} of the causal influence argument. The explanation presented 
here may of course be at most a motivation and not a stringent derivation of 
the necessity to impose this causality condition on the extremal surfaces, 
however we will find this condition to be very useful in the discussion below.

\begin{figure}[htbp]                                 
\begin{center}                                  
\includegraphics[width=0.5\linewidth]{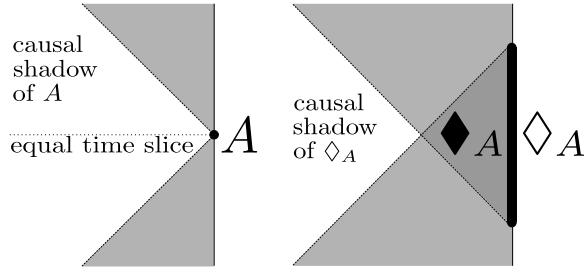}
\end{center}
\caption[]{In an RT-like prescription (left), the boundary region $A$ is 
restricted to lie on an equal time slice of the boundary, uniquely defined by a 
fixed value of the Killing time coordinate $t$. The holographic entangling 
surface belonging to $A$ lies by construction on a similarly defined equal time 
slice of the bulk spacetime, and hence in the causal shadow of $A$. The weak 
form of the causal influence argument is hence satisfied by construction. In an 
HRT-like prescription (right), $A$ can be deformed within its domain of 
dependence $\lozenge_A$ as shown in figure \ref{fig::H1}. The holographic 
entangling surface then has to be in the causal shadow of $\lozenge_A$. 
Even for a partially static spacetime where RT and HRT are expected to be 
equivalent, this strong form of the causal influence argument is a severe 
restriction, as the part of the equal time slice inside of $\blacklozenge_A$ 
would also be excluded by this argument. The region $\blacklozenge_A$ in which 
signals can be both send to and received from $\lozenge_A$ is called the 
\textit{causal wedge}, see \cite{Hubeny:2013gba}. The complement $\bar{A}$ of 
$A$ and its domain of dependence $\lozenge_{\bar{A}}$ cause a similar causal 
shadow that extends into the bulk. 
}
\label{fig::H2}
\end{figure}

This condition can still be strengthened, and to explain how let us describe
the way in which this argument comes into play in RT- and HRT-like 
prescriptions. As explained in section \ref{sec::2}, the RT-like prescription 
assumes a static bulk spacetime, in which due to the presence of a timelike 
Killing vector field one can unambiguously define a foliation of the bulk 
spacetime (as well as of the conformal boundary) into spacelike slices. Both the 
boundary region $A$ and the co-dimension two surface $\mE_A$ are then embedded 
in one of these spacelike slices by construction and the weak form of the causal 
influence argument is therefore trivially satisfied, see the left side of 
figure \ref{fig::H2}.

\begin{figure}[htbp]                                 
\begin{center}                                  
\includegraphics[width=0.6\linewidth]{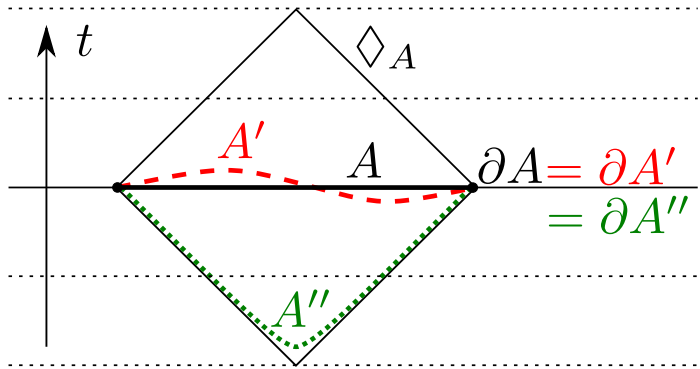}
\end{center}
\caption[]{In an HRT-like approach to entanglement entropy, the entangling 
region is not necessarily fixed to an equal time ($t=$const.) slice such as 
$A$. In fact, as long as the region stays spacelike and the boundary $\partial 
A$ stays fixed (and hence stays inside of the domain of dependence 
$\lozenge_A$), one can deform $A$ to take shapes such as $A'$ and $A''$ in the 
above picture.}
\label{fig::H1}
\end{figure}

On the other hand, the HRT prescription is intended to work for general 
spacetimes, and therefore the existence of a uniquely defined foliation of the 
bulk spacetime cannot be assumed. Similarly, the boundary region $A$ on the 
conformal boundary does not need to lie on an equal time slice of the boundary 
time. As long as the boundary $\partial A$ (to which $\mE_A$ will be anchored) 
is held fixed, the entanglement entropy of $A$, and $\mE_A$, are supposed to be 
independent of deformations of $A$ within its domain of dependence 
$\lozenge_A$, see figure \ref{fig::H1}. The \textit{domain of dependence} 
$\lozenge_A$ is defined as the set of all points on the boundary 
where every causal curve going through one of these points necessarily 
intersects $A$. Due to the possibility of deforming $A$ in the 
HRT prescription, the causal influence argument can be formulated in what we 
refer to as its \textit{strong form:}
\begin{align}
&\text{$\mE_A$ should be required to lie in the causal shadows of any 
possible spacelike deformations }
\nonumber
\\
&\text{($A',\bar{A}',A'',\bar{A}''...$) of both $A$ and $\bar{A}$ leaving the 
boundaries $\partial A$ and $\partial\bar{A}$ invariant.}
\nonumber
\end{align}
or equivalently
\begin{align}
\text{$\mE_A$ should be required to lie in the causal shadows of the 
interiors of both $\lozenge_A$ and $\lozenge_{\bar{A}}$.}
\nonumber
\end{align}
See also the right of figure \ref{fig::H2} for an illustration. 

It is very interesting to note that although RT- and HRT-like prescriptions are 
expected to agree on static spacetimes, the strong form of the causal influence 
argument is nontrivial in a static spacetime: It excludes a part of the 
spacelike slice on which $\mE_A$ would be located by construction in the RT-like 
prescription, and which in the RT-like prescription would not be excluded by any 
simple and obvious conditions. This means that the strong form of the 
causal influence argument points out a far from trivial difference between 
RT-like prescriptions and HRT-like prescriptions amended with a causality 
argument\footnote{Recall that for Einstein-Hilbert gravity, in 
\cite{Hubeny:2013gta} these causality arguments were argued to be necessary to 
make the RT and HRT prescriptions agree in certain cases.} in general 
gravitational theories of the form \eqref{HCaction}. For Einstein-Hilbert 
gravity it has been proven in \cite{Hubeny:2012wa,Hubeny:2013gba,Wall:2012uf} 
that holographic entangling surfaces anchored at the boundary can never enter 
the causal wedge of the corresponding boundary 
region.\footnote{Furthermore, in \cite{HEADRICK} a proof of the strong 
form of the causal influence argument in Einstein-Hilbert gravity is presented, 
assuming the null energy condition and some other technical details.} So this 
implies that if the strong form of the causality condition is 
satisfied in a HRT-like prescription, and if the HRT-like 
prescription (supplemented with the causality condition) is to agree with the 
RT-like prescription on static spacetimes, then the latter needs to obey 
conditions that can be derived in the Euclidean setting and 
act as a precursor to the causality argument in a Lorentzian setting.

In the previous sections \ref{sec::closedsurfaces} and \ref{sec::GBG} we 
showed 
that functionals of the form \eqref{generalfunctional} may in many cases allow 
for additional closed extremal surfaces in AdS black 
hole and global AdS spacetimes, see e.g. figure \ref{fig::1}. Due to standard 
AdS/CFT results, these surfaces can be expected to be unphysical. In the 
HRT prescription, the strong form of the causal influence argument 
presented above can be used to rule out these additional extremal surfaces, as 
we demonstrate in the remainder of this section.

\subsection*{Black hole backgrounds}

\begin{figure}[htbp]                                 
\begin{center}                                  
\includegraphics[width=0.4\linewidth]{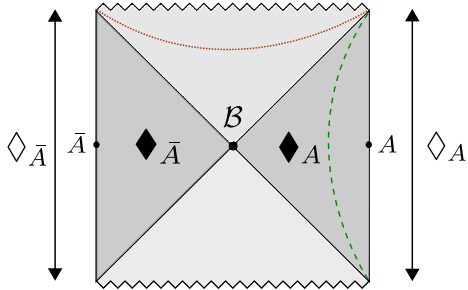}
\end{center}
\caption[]{When the region $A$ in the CFT is a full equal time slice of the 
right boundary, the causal wedges $\blacklozenge_{A}$ and $\blacklozenge_{\bar 
A}$ together cover the entire region outside the event horizons of the black and 
white holes. As illustrated by the dashed green line, all closed extremal 
surfaces outside the black hole event horizon lie in this region and are thus 
ruled out by the strong form of the causal influence argument. The dotted red 
line representing an additional closed extremal surface inside the event horizon 
is also forbidden, as it is timelike connected to $\lozenge_{A}$ and 
$\lozenge_{\bar A}$. The intersection of causal shadows of  both $\lozenge_{A}$ 
and $\lozenge_{\bar A}$ is therefore the bifurcation surface $\mathcal{B}$ of 
the black hole (which is called the causal shadow of the spacetime), 
leaving it as the only permissible closed extremal surface.}
\label{fig::ak}
\end{figure}

In black hole spacetimes, it is clear that at least when the additional 
extremal surfaces are outside of 
the black hole event horizon (case $b)$ in figure \ref{fig::1}), for a given 
time slice of the bulk spacetime there will always be an additional extremal 
surface on this slice homologous to the full boundary. In an RT-like 
prescription, the conditions on the extremal surfaces\footnote{The homology 
condition and minimisation of the entropy.} would lead one to deduce that the 
additional extremal surface outside the black hole bifurcation surface 
determines the total entropies of CFT$_{1}$ and CFT$_{2}$, since it gives a 
lower entropy. This would be a serious problem, as it would imply a 
mismatch between CFT and black hole entropy. Now let's look at this problem 
in a HRT-like (i.e. manifestly Lorentzian) framework, where causality arguments 
can be used. Consider the setup depicted in figure \ref{fig::ak}. If $A$ 
is a complete equal time slice of the right boundary 
(and hence $\bar{A}$ of the left), then $\lozenge_A$ is the complete right 
boundary in its full extent in space and time (and similarly 
$\lozenge_{\bar{A}}$ is the complete left boundary). Requiring that the 
entangling surface(s) corresponding to this division of the total system into 
subsystems $A$ and $\bar{A}$ are not connected to any point on the boundaries 
via timelike curves leaves the black hole bifurcation surface as the only 
possible extremal surface, see figure \ref{fig::ak}. Therefore, the 
strong form of the causal influence argument leads (in the cases studied 
in this paper) to an agreement between black hole entropy and holographic CFT 
entropy.

\subsection*{Global AdS}

Both for NMG (section \ref{sec::closedsurfaces}) and Gauss-Bonnet gravity 
(section \ref{sec::GBG}) we also found closed extremal surfaces in the bulk of 
global AdS spacetime. As explained in section \ref{sec::GBG}, these additional 
extremal curves have to be compared with the empty curve, which assigns zero 
entropy to the dual CFT and is allowed by the homology condition in 
topologically trivial spacetimes. As global AdS spacetime does not contain any 
event horizons, the entire bulk is in causal contact with the boundary CFT. 
Therefore, for the full boundary as CFT subsystem $A$ the causal influence 
argument rules out any curve in the bulk, leaving only the empty curve to 
correctly determine the CFT entropy as zero. This argument similarly applies to 
any other topologically trivial spacetime, such as for example boson stars 
mentioned in section \ref{sec::GBG}.

\subsection*{Curves anchored at the boundary}

Let us now come to the additional extremal curves anchored 
at the boundary that we found in section \ref{sec::anchored}, see especially 
figure \ref{fig::multiplecurves}. The calculations of that section were 
carried 
out on AdS and BTZ background spacetimes. For such background spacetimes, it is 
known that (on an equal time slice) the boundary of the causal wedge of a 
certain boundary region $A$ will be given by a spacelike geodesic of the form 
\eqref{geoAdS} and \eqref{geoBTZ} respectively, see \cite{Hubeny:2012wa}. We 
can now easily see that the additional (non-geodesic) extremal curves found 
in section \ref{sec::anchored} are ruled out by the strong form of the 
causal influence argument by examining figure \ref{fig::multiplecurves}. There, 
we see two geodesics anchored at the boundary, drawn as thick dashed (red and 
blue) lines. The first one (drawn in red) is similar to the curve labeled by 
$\mC$ in figure \ref{fig::2} and marks the boundary of the causal wedge 
$\blacklozenge_A$ corresponding to the boundary region $A$. The second one 
(drawn in blue) is the equivalent of curve $\mD$ in figure \ref{fig::2} and 
marks the boundary of $\blacklozenge_{\bar{A}}$, the causal wedge of $\bar{A}$. 
In figure \ref{fig::multiplecurves} it is now easy to see that the 
additional, non-geodesic extremal curves discussed in section 
\ref{sec::anchored} 
are excluded by the strong form of the the causal influence argument: They 
either enter $\blacklozenge_A$ or $\blacklozenge_{\bar{A}}$, and hence leave the 
causal shadow of either $A$ or $\bar{A}$. Apart from the geodesics 
which are located exactly at the boundary of the causal wedges, we didn't find 
any additional curves which are not excluded by the 
strong form of the causal influence argument.

\subsection*{Summary}

In summary, we find that for the examples of (partially) static spacetimes in 
NMG and 
Gauss-Bonnet gravity considered here, only the HRT-like prescription amended 
with the 
strong form of the causal influence argument can ensure physical results 
for entanglement entropy. This is in contrast to the Einstein-Hilbert case, 
where the RT and the HRT prescription amended with the causal influence 
argument agree on such spacetimes. For RT-like prescriptions, due to their 
manifestly Euclidean nature, it is not possible to make arguments based on 
causality. Nevertheless, if RT-like prescriptions and HRT-like prescriptions 
with causality conditions are supposed to agree on (partially) static 
spacetimes, the surfaces found in the RT setting need to obey the causality 
conditions which only make sense in a covariant framework. This implies the 
fascinating possibility that there exists a condition that has to be imposed in 
the RT-like approach in addition to the homology constraint, and which can be 
derived by arguments completely independent of causality. This condition 
would act as a precursor to the (strong form of the) causal influence argument 
that comes into effect in the Lorentzian setting. Perhaps a better understanding 
of holographic entanglement entropy would hopefully give conditions from first 
principles that exclude such pathological surfaces.

In the next section, we will investigate whether for Gauss-Bonnet 
gravity one can also derive conditions that rule out the additional 
closed extremal surfaces and fix the holographically computed CFT entropy to 
the expected physical value in the RT-like prescription. We will speculate on 
a possible connection between these conditions and the causality constraint in 
section \ref{sec::outlook}.

\section{Conical boundary condition method in Gauss-Bonnet gravity}
\label{sec::GBconstraints}

\subsection*{Method}

In \cite{Lewkowycz:2013nqa}, Lewkowycz and Maldacena proposed a method of 
calculating entanglement entropy holographically for CFT duals to 
Einstein-Hilbert gravity by introducing the concept of generalised gravitational 
entropy. This is an extension of the usual Euclidean methods for calculating 
black hole entropy to solutions without $U(1)$ symmetry in Euclidean 
time. In the present context, it is important to note that this method is 
applicable only to spatial regions in static spacetimes. To calculate 
entanglement entropy, the basic idea is to translate the replica trick into the 
bulk, where a co-dimension two hypersurface with conical defect is introduced. 
By expanding the bulk equations of motion about the conical singularity, and 
demanding finiteness of the energy-momentum tensor, the location of the 
holographic entangling surface can be determined. Thus entanglement entropy can 
be calculated. The details of this procedure for Einstein-Hilbert gravity were 
given in \cite{Lewkowycz:2013nqa}. It was also applied to Gauss-Bonnet gravity 
in \cite{Bhattacharyya:2013jma,Chen:2013qma,Bhattacharyya:2013gra}, and to 
general higher curvature theories of the form \eqref{HCaction} in 
\cite{Dong:2013qoa,Camps:2013zua}.

The metric describing the conical singularity around a co-dimension 
two bulk surface can be written as \cite{Bhattacharyya:2013gra}
\begin{align}
ds^{2} = e^{2\rho}\left(dq^2+q^2d\tau^2\right)+ 
\left(\gamma_{ij}+q\cos(\tau)k_{\left(q\right)ij}+q\sin(\tau)k_{
\left(\tau\right)ij }\right)dy^{i}dy^{j} +...
\label{expmetric}
\end{align}
with 
\begin{align}
e^{2\rho} = q^{-2\varepsilon}.
\end{align}
Here, $i=1,2,...,d-1$ and $\varepsilon$ is a small parameter that is 
later taken to zero in the replica trick. $k_{(\alpha)ij}$ ($\alpha \in 
\left\{\tau,q\right\}$) is the extrinsic curvature of the surface, and $y^i$ 
are its induced coordinates. The conical singularity is localised along the 
surface at $q = 0$.

In Einstein-Hilbert gravity, by demanding that the equations of motion near 
the singularity are satisfied and regular, one arrives at the condition 
\cite{Lewkowycz:2013nqa}
\begin{align}
k_{(\alpha)} = 0, \label{kzero}
\end{align}
i.e. the co-dimension two surface is a minimal area surface, thus proving the 
RT prescription. Note that this is only valid in static scenarios, and is 
therefore not a proof of the HRT prescription. 

For 4+1-dimensional Gauss-Bonnet gravity, upon inserting the metric 
\eqref{expmetric} the following components of the equations of motion become 
singular \cite{Bhattacharyya:2013gra}:
\begin{align}
&\text{$qq$-component:\ }&&
-\frac{\varepsilon}{q}k-\frac{\lambda 
L^2\varepsilon}{q}\Big{[}k\mR-2k_{ij}\mR^{ij}+q^{2\varepsilon}
\left(-k^3+3kk_{ij}k^{ij}-2k_{il}k^{lj}k^{i}_{j}\right)\Big{]} 
\label{qqcomp1}
\\ 
&\text{$qi$-component:\ }&&
-\frac{2\lambda 
L^2\varepsilon}{q}q^{2\varepsilon}\Big{[}k\nabla_{j}k^{j}_{i}-k\nabla_{i}
k+k^j_i\nabla_jk-k_{ij}\nabla_l k^{lj}+k_{lj}\nabla_i 
k^{lj}-k_{jl}\nabla^{j}k^{l}_{i}\Big{]} 
\label{qicomp1}
\\
&\text{$ij$-component:\ }&&
4\lambda L^2\Big{[}
\frac{\varepsilon}{q}q^{4\varepsilon}\big{(}k_{ij}k_{lm}k^{lm}-2k_{il}k^{lm}k_{
mj}+k_{
il}k^l_j k-k k_{lm}k^{lm}\gamma_{ij}+k_{ln}k^{nm}k^{l}_{m}\gamma_{ij}\big{)}
\nonumber
\\
&\ &&+\frac{\varepsilon^2}{q^2}q^{4\varepsilon}\big{(}
k^2\gamma_{ij}-2k k_{ij}-k_{lm}k^{lm}\gamma_{ij}+2k_{il}k^{l}_j\big{)}\Big{]}.
\label{ijcomp1}
\end{align}
$i,j\in\{1,2,3\}$ are the indices of the induced metric on the co-dimension two 
surface, with intrinsic Ricci tensor $\mR_{ij}$ and $k = \text{Tr}\,k_{ij}$. 
In the above equation, there is only one extrinsic curvature $k_{ij}$ as we 
are assuming a static background spacetime. In such a background, the two 
extrinsic curvatures of a surface on an equal time slice are always linear 
combinations of only one tensor $k_{ij}$, independently of the choice of normal 
coordinates. By demanding that the singular components vanish, constraints on 
the entangling surfaces in 4+1-dimensional Gauss-Bonnet gravity can be derived. 
We utilise these constraints in the following to investigate whether the closed 
extremal surfaces of the Gauss-Bonnet entropy functional \eqref{GBSEE} are valid 
in the holographic calculation of entanglement entropy.

\subsection*{Investigating conditions on closed bulk surfaces}

As shown in section \ref{sec::GBG}, assuming spherical symmetry the hypersphere 
of radius $r =\sqrt{-2\lambda}L$ on a constant time slice $t=\text{const.}$ is 
a closed extremal surface of the Gauss-Bonnet entropy functional \eqref{GBSEE} 
regardless of the topology of the bulk spacetime. That is, it is an extremal 
surface for a static spherically symmetric background of the form
\begin{align}
ds^2=-h(r)dt^2+f(r)dr^2+r^2 \gamma_{ij} dx^idx^j,
 \label{background}
\end{align}
with the line element of the hypersphere
\begin{align}
\gamma_{ij}
dx^idx^j=d\theta^2+\sin(\theta)^2d\phi^2+\sin(\theta)^2\sin(\phi)^2d\psi^2, 
\label{sphere}
\end{align}
independently of functions $f(r)$ and $h(r)$ in \eqref{background}. For a 
static black hole spacetime, the black hole bifurcation surface is also an 
extremal surface of the functional.

To subject these extremal surfaces to the constraints arising from the 
equations of motion \eqref{qqcomp1}--\eqref{ijcomp1}, the corresponding 
extrinsic curvatures need to be calculated. For a given co-dimension two 
surface characterised by embedding functions $x^{\mu}=X^{\mu}(y)$, the 
extrinsic curvature can be calculated via \cite{Chen:2013qma}\footnote{In 
contrast to the quantity $k^{(\alpha)}_{\mu\nu}$ presented in appendix 
\ref{sec::notation} where $\mu,\nu$ were indices of the full spacetime, this 
quantity is the extrinsic curvature projected to the internal space of the 
hypersurface.}
\begin{align}
k^{(\alpha)}_{ij}=-(n_{(\alpha)})_{\mu}\left(\nabla_{i}\partial_{j}X^
{ \mu }+\Gamma^{\mu}_{\rho\sigma}\partial_{i}X^{\rho}\partial_{j}X^{
\sigma}\right).
\end{align}
Here, $\mu,\rho,\sigma\in\{0,...,4\}$ are indices of the full spacetime with 
Christoffel symbols $\Gamma^{\mu}_{\rho\sigma}$, $i,j\in\{1,2,3\}$ are indices 
of the induced metric on the co-dimension two entangling surface and 
$\alpha\in\{1,2\}$ labels the two orthonormal vectors $(n_{(\alpha)})^{\mu}$ to 
this surface, see also appendix \ref{sec::notation}. 

For the extremal hyperspheres, we will use coodinates $\theta,\phi,\psi$, in 
which terms of the form $\partial_{i}X^{\mu}$ become Kronecker-deltas. By 
symmetry, the extrinsic curvature $k^{(1)}_{ij}$ with respect to the normal 
vector $n_{(1)}=\partial_t$ will vanish identically, so we are left with
\begin{align}
k_{mn} \equiv 
k^{(2)}_{mn}=-\sqrt{|f(r)|}\Gamma^{r}_{mn}=\frac{r}{\sqrt{|f(r)|}}\gamma_{mn},
\label{concretek}
\end{align}
where by slight abuse of notation we introduced ``shifted indices'' 
$m,n\in\{\theta,\phi,\psi\}$\footnote{We refer to these as 
shifted because in the latin indices the $\theta$-component 
corresponds to $i=1$, while in the greek indices the same component corresponds 
to $\mu=2$. So in the usual notation the above equation would imply 
$k_{11}\sim\Gamma^{1}_{22}$ etc.}. The singular equations of motion 
\eqref{qqcomp1}--\eqref{ijcomp1}  
simplify greatly upon insertion of \eqref{concretek}: The 
$qi$-component vanishes trivially as it only contains covariant derivatives 
$\nabla$ with respect to the induced metric $\gamma_{ij}$ and 
$k_{ij}=\text{const.}(r)\cdot\gamma_{ij}$. Using that for the hyperspheres 
$\mR_{ij}=\gamma_{ij} \mR/3$ and $\mR=6/r^2$, the 
$qq$-component simplifies to
\begin{align}
-\frac{\epsilon}{q}\left[k\left(1+\frac{2\lambda 
L^2}{r^2}\right)-\lambda L^2\frac{2}{9}q^{2\epsilon}k^3\right].
\label{rrcomp}
\end{align}
Similarly, the $ij$-component reads
\begin{align}
4\lambda 
L^2\gamma_{ij}\left[\frac{\epsilon}{q}q^{4\epsilon}k^3\frac{-2}{27}+\frac{
\epsilon^2}{q^2}q^{4\epsilon}k^2\frac{2}{9}\right].
\label{ijcomp}
\end{align}

\subsection*{Testing the conditions}

It is clear that a bifurcation surface, which necessarily is a geometrical 
extremal surface with $k=0$,\footnote{We expect \eqref{background} to have an 
event horizon where $h(r)=0=1/f(r)$, see \cite{Jacobson:2007tj} for a further 
discussion. It would then follow from \eqref{concretek} that $k=0$ at the 
bifurcation surface.} makes both \eqref{rrcomp} and \eqref{ijcomp} vanish 
identically. Additionally, we see that the leading divergence in \eqref{rrcomp} 
vanishes when the functional \eqref{GBSEE2} is extremised, i.e. for any extremal 
curve of the functional \eqref{GBSEE}. Hence the bifurcation surface of a 
black hole will always satisfy these conditions.

Let us now turn to the additional extremal surface that appears for 
example in global AdS space at radius 
$r=\sqrt{-2\lambda}L$. With extrinsic curvature \eqref{concretek}, the 
remaining 
terms in \eqref{rrcomp}, and \eqref{ijcomp} do not vanish. It was exactly the 
vanishing of this leading divergence in the $qq$-component that was shown in 
\cite{Dong:2013qoa} to be equivalent to the equations of motion derived from 
extremising the proposed entropy functional. Nevertheless, it is apparent from 
the above computations that the other components (and perhaps the less 
divergent terms) may in principle contain vital information that is 
needed to rule out the 
unphysical extremal surfaces that one finds by varying the entropy functional. 
As \eqref{qicomp1} and \eqref{ijcomp1} are proportional to $\lambda$, it is 
immediately apparent that these components do not play any role in 
Einstein-Hilbert gravity ($\lambda=0$). 

Thus, at least for the closed extremal curves in Gauss-Bonnet gravity, the 
conditions derived from the conical boundary condition method yield a 
restriction on the RT-like prescription that ensures physicality of the 
resulting entropy. As explained in section \ref{sec::Headrick}, such conditions 
are needed in order to play the role in the RT-like prescription that 
the causal influence argument plays in the HRT-like prescriptions. We 
would however like to add that although the conditions \eqref{rrcomp} and 
\eqref{ijcomp} are successful in ruling out the additional closed extremal 
curves in the RT-like prescription, there are cases 
when these constraints prove to be too restrictive. For example, it was shown 
in \cite{Bhattacharyya:2013jma} that the extremal surfaces of the entropy 
functional corresponding to cylindrical boundary regions fail to satisfy the 
conditions coming from the subleading divergent component \eqref{qqcomp1} near 
their turning point in the bulk.\footnote{We thank Aninda Sinha for 
pointing out these issues to us.} The conditions we have discussed in this 
section are therefore not yet complete for the RT-like prescription. 
Further investigation in this direction 
is worthwhile, as it seems to be a promising approach towards
 ruling out the additional curves in a Euclidean setting.

It is interesting to note that similar conditions on the entangling 
surfaces in static settings can be derived independent of the conical boundary 
condition method (i.e. the replica trick), by considering the Brown-York 
stress tensor $T$ computed on a static co-dimension one hypersurface extending 
into the bulk \cite{Bhattacharyya:2013sia,Bhattacharyya:2013jma}. The profile 
of this hypersurface that defines the extension of it into the bulk is then a 
spacelike co-dimension two hypersurface located on an equal time slice, just as 
in the calculation of entanglement entropy. Indeed, for Einstein-Hilbert 
gravity, demanding $T_{tt}=0$ results in a minimal surface condition on the 
profile of the hypersurface, in agreement with the RT prescription 
\cite{Bhattacharyya:2013sia}. This seems to imply a connection between 
entanglement entropy and the Brown-York stress tensor. In 
\cite{Bhattacharyya:2013jma} the same approach was investigated for 
Gauss-Bonnet gravity, and it was shown that demanding $T_{tt}=0$ imposes the 
following equation on the profile of the co-dimension one hypersurface:
\begin{align}
k+\lambda L^2\left(k\mR-2k_{ij}\mR^{ij}\right)+\frac{\lambda L^2}{3}
\left(-k^3+3kk_{ij}k^{ij}-2k_{il}k^{lj}k^{i}_{j}\right)=0
\label{Sinha}
\end{align}
Although not equal, this equation bears a remarkable similarity with the 
divergent part of the $qq$-component \eqref{qqcomp1} derived above from the 
conical boundary condition approach. While the first two terms in \eqref{Sinha} 
are exactly the equation of motion derived from the Jacobson-Myers functional, 
the last term corresponds to the subleading term in \eqref{qqcomp1}. This means 
that from the discussion above, the additional closed extremal surfaces fail to 
satisfy \eqref{Sinha} while only a black hole bifurcation surface would do so. 
For a spherical region on the boundary, the holographic entangling 
surfaces make the first two terms and the third term in \eqref{Sinha} vanish 
separately. Hence they are extremal curves of the Jacobson-Myers functional and 
additionally satisfy the condition of third order in extrinsic curvature that 
we needed to rule out the additional closed extremal curves in the bulk. 
However, as mentioned in the above, the second term remains too restrictive for 
other boundary regions, such as a cylinder.

For NMG, in principle it is possible to perform analyses similar to those 
presented in this section, yet this is a very tedious task. Nevertheless, at 
least for the closed extremal curves presented in section 
\ref{sec::closedsurfaces} it would 
likely yield the same results. It was already mentioned in 
\cite{Lewkowycz:2013nqa} that in cases with a full $U(1)$ symmetry in Euclidean 
time, the conical boundary conditions method reproduces Wald's entropy for 
Einstein-Hilbert gravity as well as for higher curvature theories.

\section{Discussion}
\label{sec::conclusion}

\subsection{Summary}

Let us begin this final section with a summary of this paper. We have 
investigated functional prescriptions of calculating entanglement entropy 
holographically in higher curvature gravity theories, using New Massive gravity 
(short NMG) and Gauss Bonnet gravity as concrete examples. We emphasised that 
in using entropy functionals to calculate entanglement entropy of a given 
region in a dual CFT, a non-trivial step is to find the particular surface upon 
which the corresponding functional is to be evaluated. The location of these 
surfaces can in principle be determined by solving the equations of motion with 
conical boundary conditions \cite{Lewkowycz:2013nqa}, however this is very 
involved and it is hoped that they could be alternatively determined by 
extremising the entropy functional \cite{Dong:2013qoa}. 
In this work we considered the latter approach in (partially) static 
spacetimes (i.e. AdS and black hole backgrounds), introducing the RT-like 
and HRT-like prescriptions which are based 
on extremising the entropy functional \eqref{generalfunctional} proposed for 
higher curvature theories of square order in the curvature. The RT-like 
prescription translates the Ryu and Takayanagi prescription for (partially) 
static spacetimes restricted to a constant time-slice in 
Einstein Hilbert gravity to higher curvature theories. The HRT-like prescription 
on the other hand is based on the Hubeny-Rangamani-Takayanagi prescription, 
which applies also to dynamical set ups. In both the RT- and HRT-like 
prescriptions, the extremal surfaces used to calculate CFT entropy 
holographically, for physicality, must be homologous to the full boundary. 
However in the HRT-like prescription, additional constraints can be imposed 
which arise from the Lorentzian nature of the spacetimes.

In a (partially) static scenario, the RT- and HRT-like prescriptions 
should agree, at least when a causality condition is imposed on the latter. 
To investigate this, we studied the nature of the extremal surfaces of the 
entropy functionals in NMG and Gauss-Bonnet gravity in sections 
\ref{sec::NMGtotal} and \ref{sec::GBG} respectively. In particular we focused on 
closed extremal surfaces in AdS and black hole backgrounds, whose significance 
in relation to dual CFT entropy was explained in section 
\ref{sec::closedcurves}. In black hole backgrounds, the bifurcation 
surface will always be a closed extremal surface of 
the entropy functional, however we discovered that for certain parameter ranges 
in NMG a closed extremal surface additional to the bifurcation 
surface also exists. For the non-rotating BTZ black hole background, this 
surface can encircle the black hole event horizon (see figures \ref{fig::1} and 
\ref{fig::3}), and evaluates the entropy functional at a lower value than that 
given by the black hole bifurcation surface. As explained in sections 
\ref{sec::2} and \ref{sec::closedcurves}, a naive implementation of the RT- and 
HRT-like proposals which employs only the homology constraint would hence 
require us to identify the CFT-entropy with the value given by the additional 
closed extremal surface, instead of the expected entropy.

Since NMG is plagued by several problems concerning unitarity in the bulk 
and on the boundary, in section \ref{sec::GBG} we investigated closed extremal 
surfaces for Gauss-Bonnet gravity, which is believed to be much better behaved 
and understood in a holographic context. We found that this theory also allows 
for additional closed bulk extremal surfaces, although for the rather 
unconventional choice of the Gauss-Bonnet coupling parameter $\lambda<0$ (or 
$\lambda>0$ in hyperbolic spacetimes). We found these additional extremal 
surfaces in topologically trivial spacetimes such as global AdS and boson stars. 
To us, our findings suggest that any functional of the type 
\eqref{generalfunctional} that is complicated enough can in principle, at 
least for certain choices of the 
parameters, allow for additional closed bulk extremal curves.

This would have the implication that, when naively employing the 
prescriptions for calculating holographic entanglement entropy, the phenomenon 
of a seeming mismatch between CFT entropy and bulk entropy could be quite 
common in higher curvature theories. We hence adopted the view 
that the additional bulk extremal surfaces described above are unphysical.

In section \ref{sec::Headrick} we therefore identified the strong form of 
the 
causal influence argument \cite{HEADRICK,HUBENY} as a possible way to rule out 
the additional extremal surfaces encountered in this work. Nevertheless, this 
argument is intrinsically only applicable to entanglement entropy in an HRT-like 
approach, i.e. crucially it is only applicable in a Lorentzian setting. Since 
the additional extremal surfaces found appear in both Lorentzian and Euclidean 
settings, it would certainly be desirable to find a way to rule them out in an 
RT-like approach too.

In search of a means to dismiss the additional extremal surfaces in the 
RT-like approach, in section \ref{sec::GBconstraints} for the example of 
Gauss-Bonnet gravity we turned to the alternative conical boundary 
condition method of finding entangling surfaces, based on translating the 
replica trick into the bulk \cite{Lewkowycz:2013nqa}. Although it was argued 
in \cite{Dong:2013qoa} that the equations of motion derived by extremising the 
functional \eqref{generalfunctional} are equivalent to the conditions arising 
from the conical boundary condition method, we showed that this method has 
the potential to provide further conditions that rule out the additional closed 
extremal surfaces, leaving only the black hole bifurcation surface (if present) 
as the correct answer for the surface determining the full CFT 
entropy. However, as we noted there are examples where these additional 
conditions seem to be too restrictive (e.g. a cylindrical entangling region in 
the boundary CFT). Further investigation into the application of these 
conditions would therefore be required.

\subsection{Outlook}
\label{sec::outlook}

\subsubsection*{Spacetime from entanglement}

We have given arguments throughout this work why the extremal surfaces 
additional to the black hole bifurcation surface should not be considered as 
defining the physical entanglement entropy of the CFT. Recently in 
\cite{Headrick:2013zda} a number of results about entanglement entropy in the RT 
prescription for Einstein-Hilbert gravity with matter satisfying the null 
energy condition were proven. Amongst other properties, it was shown that in 
this case, the extremal surface computing the full CFT entropy will always be 
the bifurcation surface of the bulk event horizon, if it exists. Hence the 
existence of additional extremal surfaces is excluded in this setting. Although 
there is no straightforward generalisation of many of the proofs given in 
\cite{Headrick:2013zda} to higher curvature theories and functionals of 
the form \eqref{generalfunctional}, it is interesting to note that the proof of 
the theorem mentioned above was the only one in that paper which made use of 
the equations of motion in Einstein-Hilbert gravity (and an energy condition on 
matter)\footnote{The same assumptions were made (amongst other technical 
details) in the proof of the causal influence argument presented in 
\cite{HUBENY}.}. In fact, our findings show that when varying functionals of 
the form \eqref{generalfunctional} without imposing causality constraints (as 
proposed in \cite{Dong:2013qoa}), this theorem does not generalise to higher 
curvature theories. We therefore think that this property is in fact the most 
non-trivial, and hence the physically most interesting of the results proven in 
\cite{Headrick:2013zda}. In the past, ideas have been proposed that 
the holographic entangling surface $\mE_A$ and the homology surface $\mF$ 
corresponding to a boundary region $A$ (see figure \ref{fig::RT}) define the 
part of the spacetime that can be holographically reconstructed from knowledge 
of the density matrix $\rho_A$, see 
\cite{Wall:2012uf,Czech:2012bh,Headrick:2013zda}. In the framework of this 
conjecture, the additional extremal surfaces would be unphysical, as the 
corresponding homology surface $\mF$ would not reach as deep into the bulk as 
for the bifurcation surface, meaning that even full knowledge of the CFT would 
not be enough to reconstruct the entire spacetime from the boundary down to the 
event horizon.

\subsubsection*{ER=EPR}

Another nice (but far from rigorous) argument against the validity of these 
additional extremal surfaces is the ``ER=EPR'' conjecture 
(see \cite{Maldacena:2013xja} and the ever growing list of papers citing this). 
It postulates a connection between non-traversable wormholes (ER) and 
entanglement (EPR), and is best explained for the example of an eternal black
 hole such as in figure \ref{fig::conf}. Here, the two CFTs are proposed to 
be in an entangled thermodouble state \cite{Maldacena:2001kr}, and in the bulk 
their entanglement is supposed to be described by the presence of the 
Einstein-Rosen bridge (left side of figure \ref{fig::2}). Concretely, the 
entanglement entropy between the two CFTs that can be calculated from the 
thermodouble state is equal to the bulk black hole entropy, given by the area 
of the bifurcation surface $\mB$ which is exactly the throat of the wormhole. 
If the additional closed extremal surfaces found in section 
\ref{sec::closedsurfaces} could not be ruled out, it 
would mean that the entanglement entropy between the two CFTs would no longer be 
equal to the black hole entropy. One might hope that the ER=EPR conjecture could 
at least still hold qualitatively in such cases, but even this is not true: 
While the existence of the bifurcation surface $\mB$ is intimately related to 
the presence of the wormhole and hence the topology of the spacetime, this is 
not the case for the additional closed extremal surfaces $\mA$: As we argued in 
section \ref{sec::closedcurves}, due to the flow of the 
Killing vector field $\partial_t$ the bifurcation surface $\mB$ will always be 
an extremal surface of any functional, whereas additional curves $\mA$ only 
depend on the local geometry of the spacetime and the precise form of the 
entropy functional employed. From the perspective of ER=EPR, when the full 
spacetime contains a wormhole entanglement between the two asymptotic boundaries 
is connected to spacetime topology as well as to extremal surfaces.\footnote{On 
the other hand, there do exist spacetimes where a bifurcation surface $\mB$, and 
hence a non-traversable wormhole is present without having two asymptotic CFTs 
that might be entangled in an obvious way, for example the bag-of-gold 
spacetime. See: \cite{Hubeny:2013gta,Freivogel:2005qh,Marolf:2008tx}.}
As the additional extremal closed curves don't depend on the spacetime 
topology, and can even appear for topologically trivial spacetimes, they 
are not consistent with ER=EPR. Yet, as concluded in this paper, these curves 
can be ruled out for example by the causal influence argument.

\subsubsection*{Causality and other additional conditions}

As explained in section \ref{sec::Headrick}, in a Lorentzian (HRT-like) 
setting 
the strong form of the causal influence argument elegantly rules out the 
additional extremal surfaces, at least for the examples considered in this 
paper. In that section we also pointed out that even in 
(partially) static spacetimes this argument implies that further restrictions 
should also be imposed in the RT-like approach if both are supposed to agree. It 
would hence be of great interest to better understand this argument, and whether 
it maybe is only the Lorentzian corollary of a more general restriction that has 
to be imposed both on HRT- and RT-like prescriptions. In section 
\ref{sec::GBconstraints}, we investigated whether for Gauss-Bonnet 
gravity in a Euclidean setting that such conditions might arise from an approach 
using conical boundary conditions. It would be interesting to find out whether 
such conditions for general surfaces and theories also have the effect of 
ruling out surfaces that would violate the causality condition in a Lorentzian 
setting. This would have the fascinating implication that the Euclidean 
computations already ``know'' in a sense about the causality in the Lorentzian 
setting.

\subsubsection*{$f(R)$ gravity}

Another interesting model of higher curvature theories are $f(R)$ 
theories, where the gravitational Lagrangian is an 
arbitrary function of the Ricci scalar $R$. As pointed out in 
\cite{Dong:2013qoa}, using a field redefinition these theories can be mapped to 
Einstein-Hilbert gravity minimally coupled to a scalar field, where holographic 
entanglement entropy can be studied using the area functional \eqref{RT}. 
In this Einstein frame one can then apply the results of 
\cite{Headrick:2013zda}. It 
would certainly be interesting to study whether $f(R)$ theories allow for 
additional extremal surfaces similarly to NMG and Gauss-Bonnet gravity, and 
how extremal surfaces are mapped into the Einstein frame. This might shed 
some 
light on the conditions that have to be imposed on extremal surfaces in 
higher curvature theories.

\section{Acknowledgements}
We would like to thank Yves Brihaye, Betti Hartmann, Matthew Headrick, 
Aninda Sinha and Jia-ju Zhang for 
helpful correspondence, as well as  Arpan Bhattacharyya, Ralph Blumenhagen, 
Olaf Hohm, Da-Wei Pang and 
Ivo Sachs for useful discussions. C.S. is grateful to Alejandra Castro, Jan 
Rosseel and Marika Taylor for interesting and helpful discussions. We would 
also like to thank 
Ann-Kathrin Straub, Migael Strydom and Hansj\"org Zeller for comments on the 
draft.

\appendix

\section{Notation}
\label{sec::notation}

In this section we are going to clarify some of the notation used in this 
paper, especially in equation \eqref{generalfunctional} that reads:
\begin{align}
\mS_{EE}=\frac{1}{4G_N}
\int_{\Sigma}\! d^{d-1}y \sqrt{\gamma} \,
\left[1+2aR+b\left(R_{\|}-\frac{1}{2}
k^2\right)+2c\left(R_{\| \|}-\text{Tr}(k)^2\right)\right]
\nonumber
\end{align}
Here, $\Sigma$ is a spacelike co-dimension two surface extending into the bulk. 
There are $d-1$ coordinates $y^i$ on this surface, and the induced metric 
is $\gamma_{ij}$ with determinant $\gamma>0$. 

Let us first give the definitions in case of a Euclidean bulk metric 
$g_{\mu\nu}$. Being co-dimension two, 
$\Sigma$ has two normal vectors $n_{\left(\alpha\right)}^{\mu}$ with $\alpha\in\{1,2\}$ and 
\begin{align}
n_{\left(1\right)}^{\mu}n_{\left(1\right)}^{\nu}g_{\mu\nu}=n_{\left(2\right)}^{\mu}n_{\left(2\right)}^{\nu}g_{\mu\nu}&=+1, \quad 
n_{\left(1\right)}^{\mu}n_{\left(2\right)}^{\nu}g_{\mu\nu}=0.
\end{align}
We then define the projections
\begin{align}
 R_{\|}\equiv R_{\mu\nu}n_{\left(\alpha\right)}^{\mu}n_{\left(\alpha\right)}^{\nu}, \quad
 R_{\| \|}\equiv 
R_{\mu\rho\nu\sigma}n_{\left(\alpha\right)}^{\mu}n_{\left(\alpha\right)}^{\nu}n_{\left(\beta\right)}^{\rho}n_{\left(\beta\right)}^{\sigma}
\end{align}
where double greek indices imply summation. The extrinsic 
curvature terms are defined via \cite{Fursaev:2013fta} 
\begin{align}
 h_{\mu\nu}&=g_{\mu\nu}-(n_{\left(\alpha\right)})_{\mu}(n_{\left(\alpha\right)})_{\nu}
 \label{projector}
 \\ \nonumber
 \\
 k^{\left(\alpha\right)}_{\mu\nu}&=h^{\lambda}_{\mu}h^{\rho}_{\nu}(n_{\left(\alpha\right)})_{\lambda;\rho}
 \label{extrinsic}
 \\ \nonumber
 \\ 
 k^2&=(k^{\left(\alpha\right)})^{\mu}_{\mu}(k^{\left(\alpha\right)})^{\nu}_{\nu}
 \label{ksq}
 \\ \nonumber
 \\
 \text{Tr}(k)^2&=(k^{\left(\alpha\right)})^{\mu}_{\nu}(k^{\left(\alpha\right)})^{\nu}_{\mu}.
 \label{Trksq}
\end{align}

For a Lorentzian bulk metric $g_{\mu\nu}$ these equations have to be modified 
as follows:
\begin{align}
 n_{\left(1\right)}^{\mu}n_{\left(1\right)}^{\nu}g_{\mu\nu}&=-1,\;
 n_{\left(2\right)}^{\mu}n_{\left(2\right)}^{\nu}g_{\mu\nu}=+1,\;
n_{\left(1\right)}^{\mu}n_{\left(2\right)}^{\nu}g_{\mu\nu}=0 
\\ \nonumber
\\
 R_{\|}\equiv 
R_{\mu\nu}n_{\left(\alpha\right)}^{\mu}n_{\left(\alpha\right)}^{\nu}&=-R_{\mu\nu}n_{\left(1\right)}^{\mu}n_{\left(1\right)}^{\nu}+R_{\mu\nu}n_
{\left(2\right)}^{\mu}n_{\left(2\right)}^{\nu}
\ \ (\text{similarly for $R_{\| \|}$)} 
\\ \nonumber
\\ 
h_{\mu\nu}&=g_{\mu\nu}+(n_{\left(1\right)})_{\mu}(n_{\left(1\right)})_{\nu}-(n_{\left(2\right)})_{\mu}(n_{\left(2\right)})_{\nu} 
 \\ \nonumber
 \\ 
 k^{\left(\alpha\right)}_{\mu\nu}&=h^{\lambda}_{\mu}h^{\rho}_{\nu}(n_{\left(\alpha\right)})_{\lambda;\rho} 
 \\ \nonumber
 \\ 
 k^2&=-(k^{\left(1\right)})^{\mu}_{\mu}(k^{\left(1\right)})^{\nu}_{\nu}+(k^{\left(2\right)})^{\mu}_{\mu}(k^{\left(2\right)})^{\nu}_{\nu} 
 \\ \nonumber
 \\
\text{Tr}(k)^2&=-(k^{\left(1\right)})^{\mu}_{\nu}(k^{\left(1\right)})^{\nu}_{\mu}+(k^{\left(2\right)})^{\mu}_{\nu}(k^{\left(2\right)})^{\nu}
_ {\mu}
\end{align}
so that effectively the indices in brackets (like $(...)_{(\alpha)}$) are 
contracted with a Minkowski metric. 
This ensures that $k^2$ and $\text{Tr}(k)^2$ are independent of the choice of 
the $n_{\left(\alpha\right)}^{\mu}$ as long as $n_{\left(1\right)}^{\mu}$ is the timelike and $n_{\left(2\right)}^{\mu}$ is 
the spacelike normal vector.

\section{Explicit equations of motion for the entropy functional in NMG}
\label{appendix::EOM}

In this section we will work with the (non-rotating) BTZ metric in 
Schwarzschild like coordinates:
\begin{align}
 g_{\mu\nu}=
 \left(
\begin{array}{ccc}
 -M+\frac{r^2}{\ell ^2} & 0 & 0 \\
 0 & \frac{1}{-M+\frac{r^2}{\ell ^2}} & 0 \\
 0 & 0 & r^2 \\
\end{array}
\right)
\end{align}
As we are going to assume that the holographic entangling curves lie in an 
equal time slice of Schwarzschild time ($t=$const., hence $r\geq 
\ell\sqrt{M}$), the sign of the $g_{tt}$-component and hence whether the metric 
is given in its Lorentzian of Euclidean form will not be relevant for us. The 
Lagrangian for a curve parameterised by $r=f(\phi),t=$const. then reads:
\begin{align}
\mL=\sigma\sqrt{f[\phi ]^2+\frac{f'[\phi ]^2}{-M+\frac{f[\phi ]^2}{\ell ^2}}} 
\left(1 +\frac{1-\ell^2k^2}{2 \sigma m^2 \ell 
^2}\right)
\end{align}
with the extrinsic curvature term
\begin{align}
&k^2=\frac{-1}{\ell ^2 \left(M \ell ^2 
f[\phi ]^2-f[\phi ]^4-\ell ^2 f'[\phi ]^2\right)^3}\times
\nonumber
\\
&\left(-2 M \ell ^2 f[\phi ]^4+f[\phi ]^6-2 M \ell ^4 f'[\phi 
]^2+f[\phi ]^2 \left(M^2 \ell ^4+3 \ell ^2 f'[\phi ]^2\right)+M \ell ^4 f[\phi ] 
f''[\phi ]-\ell ^2 f[\phi ]^3 f''[\phi ]\right)^2
\nonumber
\end{align}
Form this, we can derive the Euler-Lagrange equation of motion for $f(\phi)$, 
which takes the compact form
\begin{align}
0&=
-\big{(}M+10 m^2 M \ell ^2 \sigma \big{)} f^{16}+2 m^2 \sigma  
f^{18}+\big{(}1+4 m^2 \ell ^2 \sigma \big{)} f^{14} \big{(}5 M^2 
\ell ^2+3 f'^2\big{)}
+M \ell ^2 \big{(}13+8 m^2 \ell ^2 \sigma \big{)} 
\nonumber
\\
&
\times f^{13} f''
+M \ell ^2 \big{(}13+8 m^2 \ell ^2 \sigma \big{)} 
f^{13} f''-2 \big{(}1+m^2 \ell ^2 \sigma \big{)} f^{15} 
f''-60 M^3 \ell ^{10} f f'^4 f''
+M \ell ^8 
\big{(}41+2 m^2 \ell ^2 \sigma \big{)}
\nonumber
\\&
\times f f'^6 f''
-\ell ^2 f^{11} \big{(}32 M^2 \ell ^2+12 m^2 M^2 \ell ^4 \sigma +111 f'^2+6 
m^2 \ell ^2 \sigma  f'^2\big{)} f''+5 \ell ^4 f^9 f''^3-30 M^2 \ell ^{10} f f'^2 f''^3
\nonumber
\\
&
+15 \ell ^6 f^5 
\big{(}M^2 \ell ^2-2 f'^2\big{)} f''^3+20 M^2 \ell ^{10} f 
f'^3 f'' f^{(3)}-\ell ^2 f^{12} \big{(}10 M^3 \ell 
^2+20 m^2 M^3 \ell ^4 \sigma +27 M f'^2
\nonumber
\\&
+46 m^2 M \ell ^2 \sigma  f'^2-15 f''^2-20 f' f^{(3)}\big{)}+\ell ^4 
f^9 f'' \big{(}38 M^3 \ell ^2+8 m^2 M^3 \ell ^4 \sigma 
+347 M f'^2+18 m^2 M \ell ^2 \sigma  f'^2
\nonumber
\\&
+20 f' f^{(3)}\big{)}
-2 M \ell ^8 
f'^4 \big{(}5 f'^4+2 m^2 \ell ^2 \sigma  f'^4-12 M \ell ^2 
f''^2+4 M \ell ^2 f' f^{(3)}\big{)}-\ell ^4 f^7  \big{(}22 M^4 \ell ^4f''
\nonumber
\\&
+2 m^2 M^4 \ell ^6 \sigma +411 M^2 \ell ^2 
f'^2+18 m^2 M^2 \ell ^4 \sigma  f'^2-264 f'^4+6 m^2 \ell^2 
\sigma  f'^4+15 M \ell ^2 f''^2+60 M \ell ^2 f' f^{(3)}\big{)}
\nonumber
\\&
-\ell ^4 f^8 \big{(}5 M^5 \ell ^4+2 m^2 M^5 \ell ^6 \sigma 
+93 M^3 \ell ^2 f'^2+42 m^2 M^3 \ell ^4 \sigma  f'^2+237 M 
f'^4+66 m^2 M \ell ^2 \sigma  f'^4
\nonumber
\\&
-72 M^2 \ell ^2 f''^2+135 
f'^2 f''^2-96 M^2 \ell ^2 f' f^{(3)}\big{)}+f^{10} \big{(}10 M^4 \ell ^6+10 m^2 
M^4 \ell ^8 \sigma +75 M^2 \ell ^4 f'^2
\nonumber
\\&
+66 m^2 M^2 \ell ^6 \sigma  f'^2
+87 \ell ^2 f'^4+24 m^2 \ell ^4 
\sigma  f'^4-54 M \ell ^4 f''^2-72 M \ell ^4 f' 
f^{(3)}\big{)}+\ell ^6 f^5 f''\big{(}5 M^5 \ell ^4
\nonumber
\\&
+225 M^3 \ell ^2 f'^2
+6 m^2 M^3 \ell ^4 \sigma  f'^2
-525 M f'^4+12 
m^2 M \ell ^2 \sigma  f'^4+60 M^2 \ell ^2 f' f^{(3)}+20 
f'^3 f^{(3)}\big{)}+\ell ^4 f^6
\nonumber
\\&
\times \big{(}M^6 \ell ^6+54 M^4 
\ell ^4 f'^2+10 m^2 M^4 \ell ^6 \sigma  f'^2
+289 M^2 \ell ^2 f'^4+60 m^2 M^2 \ell ^4 \sigma  f'^4-175 f'^6+20 m^2 \ell 
^2 \sigma  f'^6
\nonumber
\\&
-42 M^3 \ell ^4 f''^2+342 M \ell ^2 f'^2 
f''^2-56 M^3 \ell ^4 f' f^{(3)}
-4 M \ell ^2 f'^3 
f^{(3)}\big{)}+\ell ^6 f^2 f'^2 \big{(}46 M^4 \ell ^4 f'^2
-89 M^2 \ell ^2 f'^4
\nonumber
\\&
+14 m^2 M^2 \ell ^4 \sigma  f'^4+21 
f'^6+6 m^2 \ell ^2 \sigma  f'^6+72 M^3 \ell ^4 f''^2
-84 M \ell ^2 f'^2 f''^2-4 M^3 \ell ^4 f' f^{(3)}
\nonumber
\\&
+28 M \ell ^2 f'^3 f^{(3)}\big{)}
-\ell ^6 f^3 f'' \big{(}50 M^4 \ell ^4 f'^2
-321 M^2 \ell ^2 f'^4+6 m^2 M^2 \ell ^4 \sigma f'^4
+47 f'^6+2 m^2 \ell ^2 \sigma  f'^6
\nonumber
\\&
+5 M^3 \ell ^4 f''^2-60 M \ell ^2 f'^2 f''^2
+20 M^3 \ell ^4 f' 
f^{(3)}+40 M \ell ^2 f'^3 f^{(3)}\big{)}-\ell ^6 f^4 
\big{(}12 M^5 \ell ^4 f'^2
+185 M^3 \ell ^2 f'^4
\nonumber
\\&
+18 m^2 M^3 \ell ^4 \sigma  f'^4-259 M f'^6
+34 m^2 M \ell ^2 \sigma  f'^6-9 
M^4 \ell ^4 f''^2+279 M^2 \ell ^2 f'^2 f''^2-60 f'^4 f''^2
\nonumber
\\&
-12 M^4 \ell ^4 f' f^{(3)}-8 M^2 \ell ^2 f'^3 
f^{(3)}+20 f'^5 f^{(3)}\big{)}
-2 \ell ^2 f \big{(}M^2 \ell ^4 f^2
-2 M \ell ^2 f^4+f^6-M \ell ^4 f'^2
\nonumber
\\&
+\ell ^2 f^2 f'^2\big{)}^2 f^{(4)}
\label{fourthorder}
\end{align}

\section{Closed extremal curves of other black hole solutions in NMG}
\label{appendix::BH}


For the more general rotating BTZ black hole we find through 
calculations similar to those done in section \ref{sec::NMGtotal} (best 
performed in Eddington-Finkelstein 
coordinates \cite{Chan:1994rs}) additional extremal curves at radial 
distances
\begin{align}
 r_a^2=
\frac{M}{4\sigma m^2} \pm\frac{1}{4} \sqrt{\frac{M^2-6 J^2\sigma 
m^2}{m^4}}
\label{solrotBTZ}
\end{align}
for $\sigma m^2<0$, or when $\sigma m^2>0$ and 
$|J|<\sqrt{\frac{M^2}{6\sigma m^2}}$. As $r^2>0$, it 
depends on the values and signs of $M,J,\sigma,m^2$ which of the branches in 
\eqref{solrotBTZ} gives a valid solution.


One can also look at Lifshitz black holes 
\cite{AyonBeato:2009nh} 
\begin{align}
 g_{\mu\nu}=
 \left(
\begin{array}{ccc}
 -\left(1-\frac{M\ell^2}{r^2}\right) \frac{r^6}{\ell^6} & 0 & 0 \\
 0 & \frac{1}{\frac{r^2}{\ell^2}-M} & 0 \\
 0 & 0 & r^2 \\
\end{array}
\right).
\label{Lifshitz}
\end{align}
which are solutions to NMG for $\ell^2m^2=-\frac{1}{2}$ and $\sigma=+1$. 
In this metric, additional extremal curves appear only for $M<0$.


\providecommand{\href}[2]{#2}\begingroup\raggedright\endgroup

\end{document}